\documentclass{article}

\usepackage{arxiv_published}

\usepackage[utf8]{inputenc} 
\usepackage[T1]{fontenc}    
\usepackage{hyperref}       
\usepackage{url}            
\usepackage{booktabs}       
\usepackage{amsfonts}       
\usepackage{nicefrac}       
\usepackage{microtype}      
\usepackage{lipsum}
\usepackage{graphicx}

\usepackage{amsmath}
\usepackage{comment}
\usepackage{multirow}
\usepackage{makecell}
\usepackage{booktabs}
\usepackage{rotating}
\usepackage{dirtytalk}
\usepackage{xcolor}
\usepackage{hyperref}
\usepackage{array}
\usepackage{arydshln}
\usepackage{algpseudocode}
\usepackage{algorithm}

\title{FedSynthCT-Brain: A Federated Learning Framework for Multi-Institutional Brain MRI-to-CT Synthesis}

\author{
  Ciro Benito Raggio\thanks{Corresponding author} \\
  Institute of Biomedical Engineering\\
  Karlsruhe Institute of Technology\\
  Fritz-Haber-Weg 1, Karlsruhe 76131\\
  Baden-Württemberg, Germany \\
  \texttt{ciro.raggio@kit.edu} \\
   \And
  Mathias Krohmer Zabaleta \\
  Institute of Biomedical Engineering\\
  Karlsruhe Institute of Technology\\
  Fritz-Haber-Weg 1, Karlsruhe 76131\\
  Baden-Württemberg, Germany
  \And
  Nils Skupien \\
  Institute of Biomedical Engineering\\
  Karlsruhe Institute of Technology\\
  Fritz-Haber-Weg 1, Karlsruhe 76131\\
  Baden-Württemberg, Germany
  \And
  Oliver Blanck \\
  Department of Radiation Oncology \\
  University Medical Center Schleswig-Holstein\\
  Feldstrasse 21, Kiel 24105\\
  Schleswig-Holstein, Germany
  \And
  Francesco Cicone \\
  Department of Experimental and Clinical Medicine\\
  Magna Graecia University\\
  Viale Europa, Catanzaro 88100\\
  Italy
  \And
  Giuseppe Lucio Cascini \\
  Department of Experimental and Clinical Medicine\\
  Magna Graecia University\\
  Viale Europa, Catanzaro 88100\\
  Italy
  \And
  Paolo Zaffino \\
  Department of Experimental and Clinical Medicine\\
  Magna Graecia University\\
  Viale Europa, Catanzaro 88100\\
  Italy
  \And
  Lucia Migliorelli \\
  Department of Information Engineering\\
  Università Politecnica delle Marche\\
  Via Brecce Bianche 12, Ancona 60131\\
  Italy
  \And
  Maria Francesca Spadea \\
  Institute of Biomedical Engineering\\
  Karlsruhe Institute of Technology\\
  Fritz-Haber-Weg 1, Karlsruhe 76131\\
  Baden-Württemberg, Germany
}

\begin{document}
\maketitle

\newpage

\begin{abstract}
The generation of Synthetic Computed Tomography (sCT) images has become a pivotal methodology in modern clinical practice, particularly in the context of Radiotherapy (RT) treatment planning. The use of sCT enables the calculation of doses, pushing towards Magnetic Resonance Imaging (MRI) guided radiotherapy treatments. Moreover, with the introduction of MRI-Positron Emission Tomography (PET) hybrid scanners, the derivation of sCT from MRI can improve the attenuation correction of PET images.

Deep learning methods for MRI-to-sCT have shown promising results, but their reliance on single-centre training dataset limits generalisation capabilities to diverse clinical settings. Moreover, creating centralised multi-centre datasets may pose privacy concerns. To address the aforementioned issues, we introduced FedSynthCT-Brain, an approach based on the Federated Learning (FL) paradigm for MRI-to-sCT in brain imaging. This is among the first applications of FL for MRI-to-sCT, employing a cross-silo horizontal FL approach that allows multiple centres to collaboratively train a U-Net-based deep learning model. We validated our method using real multicentre data from four European and American centres, simulating heterogeneous scanner types and acquisition modalities, and tested its performance on an independent dataset from a centre outside the federation.

In the case of the unseen centre, the federated model achieved a median Mean Absolute Error (MAE) of $102.0$ HU across 23 patients, with an interquartile range of $96.7-110.5$ HU. The median (interquartile range) for the Structural Similarity Index (SSIM) and the Peak Signal to Noise Ratio (PNSR) were  $0.89 (0.86-0.89)$ and  $26.58 (25.52-27.42)$, respectively.

The analysis of the results showed acceptable performances of the federated approach, thus highlighting the potential of FL to enhance MRI-to-sCT to improve generalisability and advancing safe and equitable clinical applications  while fostering collaboration and preserving data privacy.

\end{abstract}

\keywords{Federated Learning \and Medical Image Synthesis \and Synthetic Computed Tomography \and Deep Learning \and Image-To-Image Translation \and Data Sharing \and Data Privacy}

\section{Introduction}
\label{introduction}
Intermodality medical Image-To-Image (I2I) translation refers to the process of mapping images from one modality (e.g., Magnetic Resonance Imaging - MRI, Computed Tomography - CT, Positron Emission Tomography - PET) to another, while preserving essential structural and anatomical information. Deep Learning (DL) based techniques have been shown to outperform traditional methods based on image registration and/or voxel-wise analysis \cite{DAYARATHNA2024} and they have gained attention in several clinical fields, the most important being the derivation of Synthetic CT (sCT) from MRI or Cone Beam CT (CBCT), to facilitate radiotherapy planning/adaptation or to improve attenuation correction in PET/MRI acquisition \cite{Spadea2021, Thummerer2024, Boulanger2021}.

Despite the advancements in DL-based I2I, several critical limitations remain. Indeed, current DL approaches are trained on datasets that represent specific patient populations, which hampers their generalizability across different clinical settings \cite{LIU2021101953, Graf2023, Ozbey2023, Bahloul_Jabeen_2024}. For instance, the training datasets often exclude patients with implants or atypical anatomy, resulting in synthetic CTs that are less reliable for these unique cases \cite{Grigo2024}. Additionally, single-site models tend to underperform when applied to images from different medical centres, as they are sensitive to domain shifts in imaging protocols, scanner types, and patient demographics \cite{Dalmaz2024}. 

To improve generalisability in MRI-to-sCT and foster innovation in the field, several studies have been conducted using cross-centre and multi-centre datasets to create models that generalise to prostate and pelvic images from different sources \cite{Tahri2023, Texier2023, Han2025}. The study by Han et al. \cite{Han2025} used data from multiple centres also to address the impact of pre-processing to harmonise images and reduce the variability between images using normalisation, clipping and intensity correction techniques. Additionally, it investigated the impact of various combinations of data augmentation techniques for the MRI-to-sCT task focusing on deformation and rotation techniques in the case of multiple centres. 

Furthermore, initiatives like the SynthRAD2023 \cite{huijben2024generating} challenge have been proposed by the scientific community. These efforts allow DL models — such as the MSEP framework, the challenge winner \cite{Huijben2024, chen2023hybrid} — to be trained on multi-centre datasets using both brain and pelvic regions, providing access to more diverse patient data. While directly pooling medical images from multiple sites could intuitively address the issue of small sample sizes, this approach faces significant barriers due to stringent privacy protection policies such as the Health Insurance Portability and Accountability Act (HIPAA) \footnote{\url{https://www.hhs.gov/hipaa/index.html}} in the United States and the General Data Protection Regulation (GDPR)\footnote{\url{https://gdpr-info.eu/}} in Europe. These regulations impose limitations on the exchange of personal health data, posing practical challenging to sharing and centralising medical images across sites \cite{kaissis2020secure, guan2024federated}. 

The presence of bias from limited diversity in images from a single institution, along with concerns about data centralisation, underscores the need for a collaborative approach that enables access to substantial amounts of data without centralising it \cite{adnan2022federated}. 

Federated Learning (FL) can overcome data-sharing barriers by allowing multiple institutions to train a DL model collaboratively without sharing their own data on a central server. This approach both facilitates secure data collaboration and also leverages distributed computing resources to strengthen model generalizability across diverse, unseen datasets \cite{HernandezCruz2024}. 

Given these premises, the aim of the proposed study was to establish a FL framework for MRI-to-sCT, which we call FedSynthCT, and to prove its validity in the context of brain imaging (henceforth we will refer to our approach as FedSynthCT-Brain). 

A benchmark analysis was conducted across multiple architectures and different aggregation strategies in order to assess the trade-offs between efficiency and effectivenesses for realistic clinical applications.

To the best of the authors' knowledge, this is among the first applications in the literature to employ a FL paradigm in MRI-to-sCT. The main contributions and innovative aspects of our work are:

\begin{itemize}
    \item The proposal of a cross-silo horizontal FL approach for MRI-to-sCT in brain imaging, where data of the same type (horizontal FL) from various centres (or \say{silos})is collaboratively used to train a deep learning model. This innovation addresses key privacy concerns by avoiding the need to pool sensitive patient data.
    \item The validation of the proposed approach on real multi-centre data, simulating an authentic scenario where each centre handles heterogeneous data due to differences in scanners and acquisition modalities.
    \item The testing of the proposed FedSynthCT-Brain on an independent dataset from a centre outside the federation to assess the federated model's generalization capability.
\end{itemize}

The structure of this paper is as follows: Section \ref{sec:related_works} reviews existing research on computer-aided MRI-to-sCT and the use of FL for the task of medical-image analysis. Section \ref{sec:met} outlines the dataset and the proposed FedSynthCT-Brain approach. The experimental setup is detailed in Section \ref{sec:expprot}. Section \ref{sec:res} presents the results, with further discussion in Section \ref{sec:disc}. Finally, Section \ref{sec:conc} summarizes the main findings and the impact of the proposed research while proposing future directions.

\section{Related work} \label{sec:related_works}

\subsection{Centralised MRI-to-sCT} \label{sec:centralised_mri_sct}

Over the past years, numerous DL methods have been implemented for MRI-to-sCT.
The methods and models employed differed depending on the target anatomical regions and the computational resources available. Models trained on single image planes (axial, sagittal, or coronal) were termed 2D models. The combination of independently trained 2D models on different planes during or after inference led to improvements in accuracy and was referred to as 2D+ models \cite{Spadea2021, Boulanger2021, DAYARATHNA2024}. \newline

Multi-plane models (Multi-2D) employed images from multiple planes to train a single model, leveraging inter-planar information to outperform 2D and 2D+ configurations. Consecutive slices also formed multi-channel input during 2D training, an approach known as 2.5D, which enhanced spatial coherence without requiring full 3D volumes. Smaller image segments were employed across 2D, 2D+, Multi-2D, and 2.5D methods using 2D patching \cite{Spadea2021, Boulanger2021, DAYARATHNA2024}.

More recently, the employment of entire 3D volumes or sub-volumes (also named 3D patches) as input provided the most comprehensive spatial information and mitigated inter-slice artefacts observed in 2D-based methods \cite{DAYARATHNA2024}. However, the primary limitation of 3D models lies in their higher computational demand, which requires splitting volumetric data into smaller 3D patches for processing \cite{Spadea2021}. This patch-based processing may lead to the loss of global context and structural coherence, especially when patches are analysed in isolation \cite{Lei2019b, Koh2022}. Moreover, the scalability of 3D approaches in FL settings is further constrained due to the increased communication overhead and memory requirements associated with volumetric data. These factors make 3D-DL approaches less suitable for FL real-world scenarios where limited computational resources requirements must be accounted for \cite{antunes2022federated}. \newline

The proposed model architectures and methodologies have become increasingly complex in order to achieve accurate results. Among the most widely used models are: U-Nets, Generative Adversarial Networks (GANs), Transformers, and Diffusion-based architectures \cite{Spadea2021, Boulanger2021, DAYARATHNA2024}. In addition to fundamental architectures, recent advancements in the field, such as I2I-Mamba, have demonstrated strong potential for multi-modal medical image synthesis and medical image reconstruction by leveraging state-space operators to enhance contextual interactions across different spatial orientations \cite{atli2024i2imambamultimodalmedicalimage, kabas2024physicsdrivenautoregressivestatespace}. Additionally, novel diffusion-based techniques, such as SelfRDB, have been proposed to improve reliability in multi-modal medical image translation, integrating self-consistent recursive sampling and conditional guidance to enhance information transfer between modalities \cite{arslan2024selfconsistentrecursivediffusionbridge}.

A further macro-classification of methodologies can be made by dividing supervised and unsupervised methods for synthetic CTs generation. Supervised techniques require accurate MRI and CT registration, whereas unsupervised techniques do not require registration, but have often demonstrated instability in convergence \cite{Saxena2021, Texier2024}. The study by Texier et al. \cite{Texier2024} addressed and solved this problem related to unsupervised learning, training a patch-based 3D cGAN using the \say{Content and Style Representation for Enhanced Perceptual synthesis} (CREPs) loss function and multi-centre prostate data.

Several works employed three different 2-dimensional (2D) fully Convolutional Neural Networks (CNN), to retain structural information in sCTs while simultaneously reducing computational costs, especially when using U-Nets \cite{DAYARATHNA2024, Li2018, SpadeaDCNN2019, Hsu2022}. These 2D methods, by processing slices independently or in parallel, inherently simplify training in distributed environments, such as FL, where data sharing is constrained by computational heterogeneity \cite{antunes2022federated}.\newline

\subsection{Federated learning in medical imaging}

The growing number of studies applying FL in healthcare, particularly in medical imaging \cite{guan2024federated, Sheller2019-jo, Wenqi_Li2019, Daiqing_Li2020, Chang2020}, reflects its increasing importance. By keeping sensitive information, such as patient data, within local environments (e.g., clinical centres), FL addresses the crucial need for confidentiality in medical applications, enabling collaborative large-scale model development without data sharing. 

In the medical context, the focus is on cross-silo FL \cite{Xu_2022_CVPR}, characterised by a limited number of participating entities (often ranging from 2 to 100), where each entity is recognised as a \say{silo} of data. During a single round --a single communication cycle between the aggregation server and a set of clients (or silos)--, clients perform local training of their models using private data on the device itself, and then share updates with the server for aggregation.

The partitioning of data is of pivotal importance in this context, particularly in the domain of medical imaging, where datasets are frequently scarce, heterogeneous, and non-Independent and Identically Distributed (non-IID). Data imbalance has the potential to significantly impact model generalisation and performance across institutions, as medical images from disparate sources are susceptible to bias due to differences in equipment, protocols, and labelling processes. FL addresses these challenges by aggregating models from various institutions, thereby mitigating bias and improving generalisability \cite{Sandhu2023}.  

Several studies investigated the use of FL on medical images, especially for classification and segmentation tasks. 

In the field of classification, FL has been applied X-ray and CT images for the diagnosis of COVID-19 \cite{Feki2021, Yang2021}. Furthermore, numerous studies have focused also on the classification of skin diseases and lesions \cite{Wicaksana2022, Hossen2023}, prostate classification \cite{Yan2021, Wicaksana2022}, as well as in the ophthalmology field for the diabetic retinopathy classification \cite{JulianAndTimothy2021}.

More recently, FL approaches were proposed for segmentation tasks, such as for brain tumour \cite{Tuladhar2022, Albalawi2024}, prostate \cite{Zhu2023}, optic cup and disc segmentation \cite{Qiu2023, Wang2024}. This led to the development of frameworks that integrate the aforementioned works and are capable of performing segmentations of multiple anatomical structures or different image modalities using FL and foundational models, such as the FedFMS \cite{Liu2024}, which was created to integrate the Segment Anything Model (SAM) \cite{Kirillov_2023_ICCV} into the medical FL context.

Recent advancements in medical image synthesis have introduced innovative methods, such as FedMed-GAN, which specifically addresses the challenge of synthesizing cross-modality brain MRI images in an FL setting \cite{Wang2023}. It employs Federated Averaging (FedAvg), the \textit{de facto} aggregation strategy algorithm for FL \cite{Xu_2022_CVPR}, enabling the combination of locally trained models from multiple institutions by computing a weighted average of their parameters. 

Inspired by the FL research, this study investigates the feasibility of FL for MRI-to-CT synthesis, addressing a critical gap in the literature. Indeed, to the best of the authors' knowledge, this is among the first work to apply a cross-silo horizontal FL approach in this context, leveraging multi-center real heterogeneous datasets.

\section{Methods}
\label{sec:met}

\subsection{Datasets and preprocessing}
\label{subsec:data_preparation}

\begin{table}[tbp]
\centering
\footnotesize
\setlength{\tabcolsep}{0.005\linewidth}  
\begin{tabular}{ccccccc}
\toprule
\textbf{Parameter} &
   \multicolumn{1}{c}{} &
  \textbf{Centre A} &
  \textbf{Centre B} &
  \textbf{Centre C} &
  \textbf{Centre D} &
  \textbf{Centre E} \\ \midrule
\textbf{Patients \#} &
   \multicolumn{1}{c}{} &
  15 &
  14 &
  21 &
  29 &
  23 \\ \midrule
\multirow{2}{*}{\textbf{\begin{tabular}[c]{@{}c@{}}Voxel size \\ {[}mm$^{3}${]}\end{tabular}}} &
  \textbf{MR} &
  1,1,1 &
  1,1,1 &
  0.78,0.78,1 &
  \begin{tabular}[c]{@{}c@{}}0.98-1.12,\\ 0.98-1.12,\\ 0.98-1.12\end{tabular} &
  0.98,0.98,0.98 \\ \cmidrule(l){2-7} 
 &
  \textbf{CT} &
  \begin{tabular}[c]{@{}c@{}}0.49-0.67,\\ 0.49-0.67,\\ 2.5\end{tabular} &
  \begin{tabular}[c]{@{}c@{}}0.98,\\ 0.98,\\ 3.27\end{tabular} &
  \begin{tabular}[c]{@{}c@{}}0.78,\\ 0.78,\\ 1\end{tabular} &
  \begin{tabular}[c]{@{}c@{}}0.69-0.78,\\ 0.69-0.79,\\ 1-3\end{tabular} &
  \begin{tabular}[c]{@{}c@{}}0.59-1.27,\\ 0.59-1.27,\\ 1-2\end{tabular} \\ \midrule
\multirow{2}{*}{\textbf{Scanners}} &
  \textbf{\begin{tabular}[c]{@{}c@{}}MR \\ {[}T{]}\end{tabular}} &
  \begin{tabular}[c]{@{}c@{}}MAGNETOM \\ Trio\\ {[}3{]}\end{tabular} &
  \begin{tabular}[c]{@{}c@{}}Biograph \\ mMR\\ {[}3{]}\end{tabular} &
  \begin{tabular}[c]{@{}c@{}}Vantage \\ Titan\\ {[}1.5{]}\end{tabular} &
  \begin{tabular}[c]{@{}c@{}}MAGNETOM \\ Avanto\_fit, \\ Skyra, \\ Vida\_fit, \\ Prisma\_fit\\ {[}1.5-3{]}\end{tabular} &
  \begin{tabular}[c]{@{}c@{}}MAGNETOM\\ Aera, \\ Avanto\_fit\\ {[}1.5-3{]}\end{tabular} \\ \cmidrule(l){2-7} 
 &
  \textbf{\begin{tabular}[c]{@{}c@{}}CT\\ {[}kVp{]}\end{tabular}} &
  \begin{tabular}[c]{@{}c@{}}LightSpeed \\ QX/i\\ {[}140{]}\end{tabular} &
  \begin{tabular}[c]{@{}c@{}}Discovery \\ ST\\ {[}120{]}\end{tabular} &
  \begin{tabular}[c]{@{}c@{}}Brilliance \\ Sensation \\ Open\\ {[}120{]}\end{tabular} &
  \begin{tabular}[c]{@{}c@{}}Brilliance \\ Big \\ Bore\\ {[}120{]}\end{tabular} &
  \begin{tabular}[c]{@{}c@{}}SOMATOM\\ Definition \\ AS\\ {[}120{]}\end{tabular} \\ \midrule
\multirow{3}{*}{\textbf{Size}} &
  \textbf{Axial} &
  256 &
  256 &
  226-357 &
  167-213 &
  167-262 \\ \cmidrule(l){2-7} 
 &
  \textbf{Sagittal} &
  176 &
  176 &
  512 &
  216-262 &
  200-225 \\ \cmidrule(l){2-7} 
 &
  \textbf{Coronal} &
  256 &
  248 &
  512 &
  250-277 &
  225-248 \\ \bottomrule
\end{tabular}
\caption{Summary of the most relevant characteristics of the image datasets used in the federated learning (FL) framework. Data from Centre A, B, C, and D were used to train the federated model; thus Centre A, B, C, D acted as clients. The data from the Centre E (SynthRAD Centre B) served as an independent dataset for testing the generalisability of the federated model on unseen data. For each dataset we list the number of patients, image sizes, magnetic resonance imaging (MRI) and computed tomography (CT) scanners used, and voxel spacing in millimeters.}
\label{tab:datasets}
\end{table}

\begin{figure}[tbp]
    \centering
    \includegraphics[width = 1\linewidth]{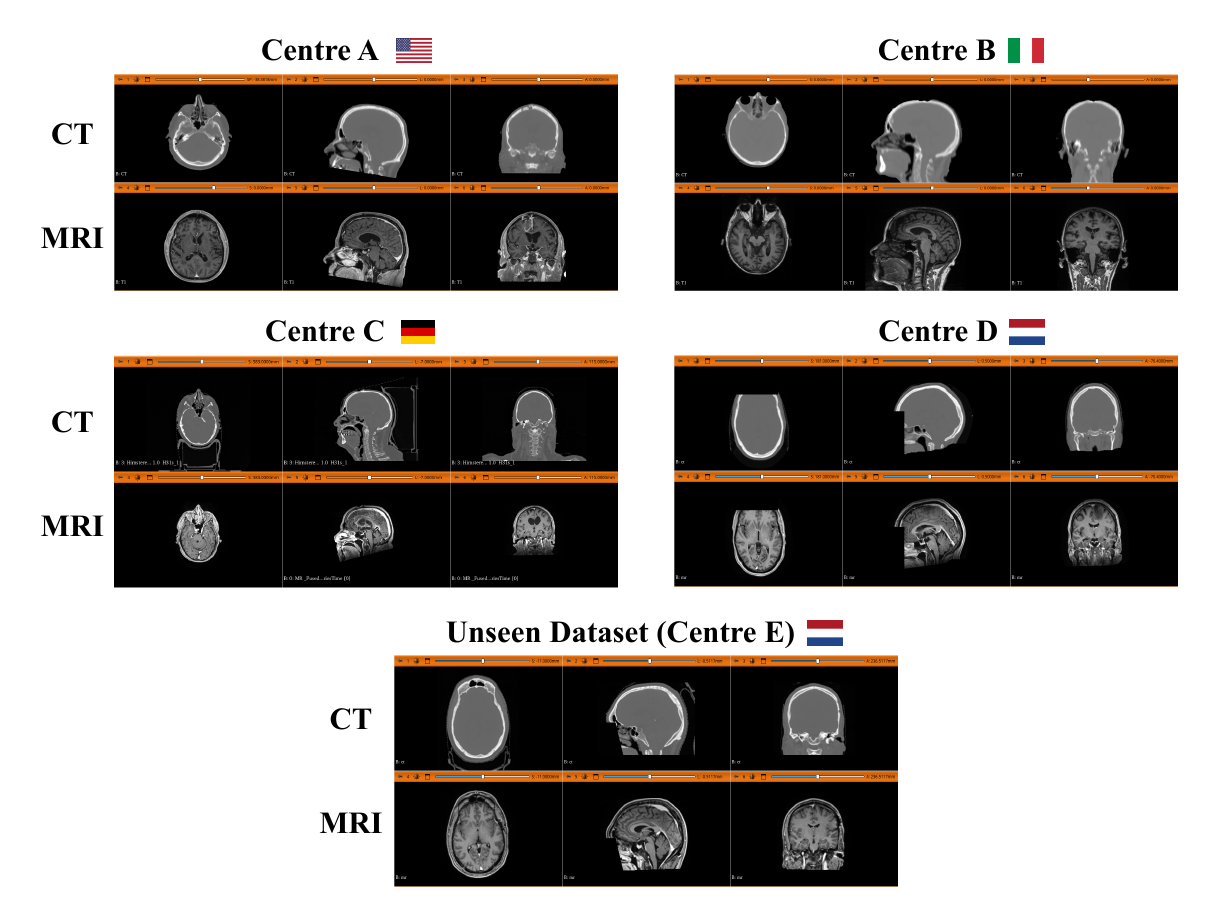}

    \caption{Visual comparison of image appearance across datasets from multiple Centres. The FOV of the images from Centre A was limited to the level of the mouth and included partial cut-off of the ears. Images from centre B were acquired using a PET-MRI hybrid scanner. Furthermore the higher CT slice thickness introduced partial volume artefacts in the skull bone reconstruction. Images coming from centre C exhibited higher spatial resolution in the cranial caudal direction, but presented other problems, including partial cropping at the top of the skull and metal artefacts. Centres D and E--derived from SynthRAD2023 challenge--shared similar imaging regions with less bias fields compared to Centre C.}
    \label{fig:data_diff}
\end{figure}

For each centre (named A, B, C, D and E) a dataset containing paired, co-registered MRI T1-weighted and CT images was available. Image acquisition was performed in accordance with the ethical standards of the 1964 Declaration of Helsinki and later amendments. Written informed consent to use data for research purposes was obtained from all patients. The image acquisition protocol and scanners differed among the various centres, as it is reported in Table \ref{tab:datasets}. This led to image variability among the federated dataset and the most relevant difference can be appreciated in Figure \ref{fig:data_diff}. 

The field of view (FOV) of images from Centre A  started at the mouth region with a straight cut off and spans towards the top of the skull. A fraction of the ears was also cut off. 

MRIs from Centre B were acquired using a PET-MRI hybrid scanner. Differently from the other centres, this acquisitions were not performed for radiotherapy planning purposes. Thus, the tube energy in the CT was 120kV and the slice thickness was 3.75 mm which introduced partial volume artefacts in the upper part of the skull bone. In light of the substantial difference in the acquisition protocol and scanners, Centre B dataset was used to increase the image variability within the federation (i.e. to introduce noise) to test the robustness of the federated model and simulate a real context. 

Images belonging to Centre C presented higher spatial resolution in the cranial caudal direction compared to the other datasets. However, the Centre C's dataset presented a number of additional challenges, including: the absence of the top of the skull in some patients, the presence of the scanner bed and pillow for each CT, metal artefacts, and high bias field artefact. 

The Centre D's and Centre E's datasets derived exclusively from two different institutions of the SynthRAD2023 multi-centre dataset (SynthRAD Centre C and SynthRAD Centre B respectively), which was created for the SynthRAD challenge within different MRI-to-sCT algorithms \cite{huijben2024generating}. Before publication, the datasets were cropped to remove the facial region. The FOV covered by both imaging modalities was similar, except for a cut-off at the lower end with a slight angle in addition to a straight horizontal line. In general, Center D and E presented less bias field artefact when compared to Center C. However, between Centre D and E, it was more pronounced in the Centre E dataset. Furthermore, both Centre MRIs were obtained using gadolinium-based contrast agents, thus introducing further heterogeneity within the federation.

To address the image variability, a pre-processing strategy was implemented at each client site with the aim of harmonising data within the federation, without the exchange of data. Following \cite{Han2025}, N4 Bias Field Correction \cite{Tustison2010} was applied to the MRI data to correct intensity inhomogeneity, which can result from factors such as magnetic field variations or patient positioning. Additionally, Min-Max Normalization was applied within each client to reduce variability in pixel intensity values.

Due to the differences in acquisition protocols, the datasets included images of varying dimensions (See Table \ref{tab:datasets}). To standardize input for the federated model, all image volumes were cropped, resized, and padded to a uniform size of $256\times256\times256$ voxels (See Section \ref{sec:expprot}). This process enabled each client to preprocess its data independently while maintaining a consistent input format across the federation, thereby allowing seamless aggregation of the DL model updated from each client. 

\begin{figure}[tbp]
    \centering
    \includegraphics[width = 1\linewidth]{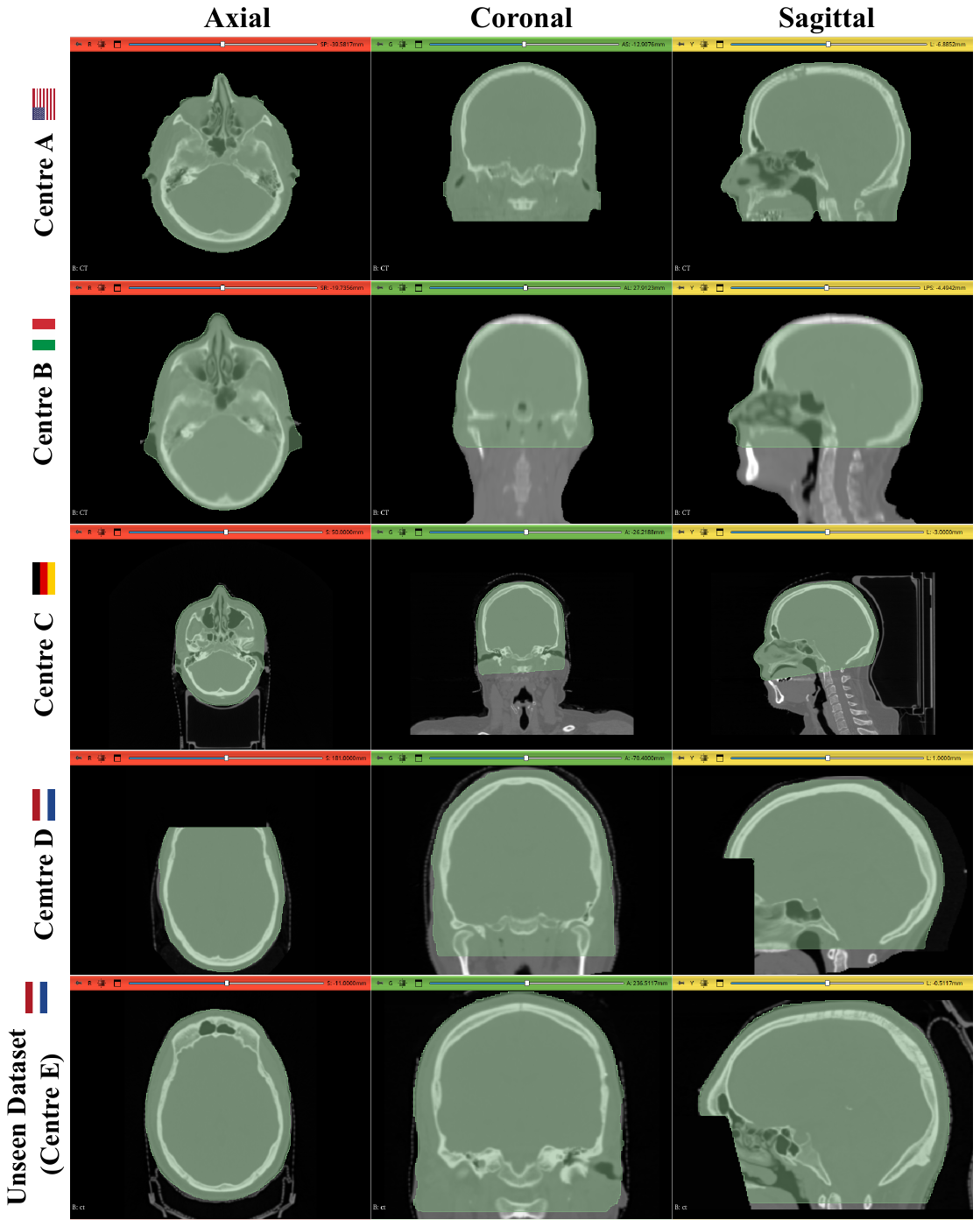}

    \caption{The figure illustrates the background masking for each institution. By eliminating potential misleading features for the model, the masking approach limits the prediction of the model to areas of interest and facilitates the harmonisation of the model's knowledge within the federation. The masking process was conducted within the individual Centre. The parts of interest were established beforehand, as would be done in a real-world context.}
    \label{fig:masking_diff}
\end{figure}

The elements outside the patient body (couch, immobilisation system, pillow) were masked out on the CT volume using 3D Slicer image computing platform \cite{Federov2012}(Figure \ref{fig:masking_diff}). The obtained mask was applied also to the corresponding MRI to remove background noise. Voxels outside the mask were set to -1000 and 0 for CT and MRI respectively.

\subsection{Federation setup}
\begin{figure}[tbp]
    \centering
    \includegraphics[width = 1\linewidth]{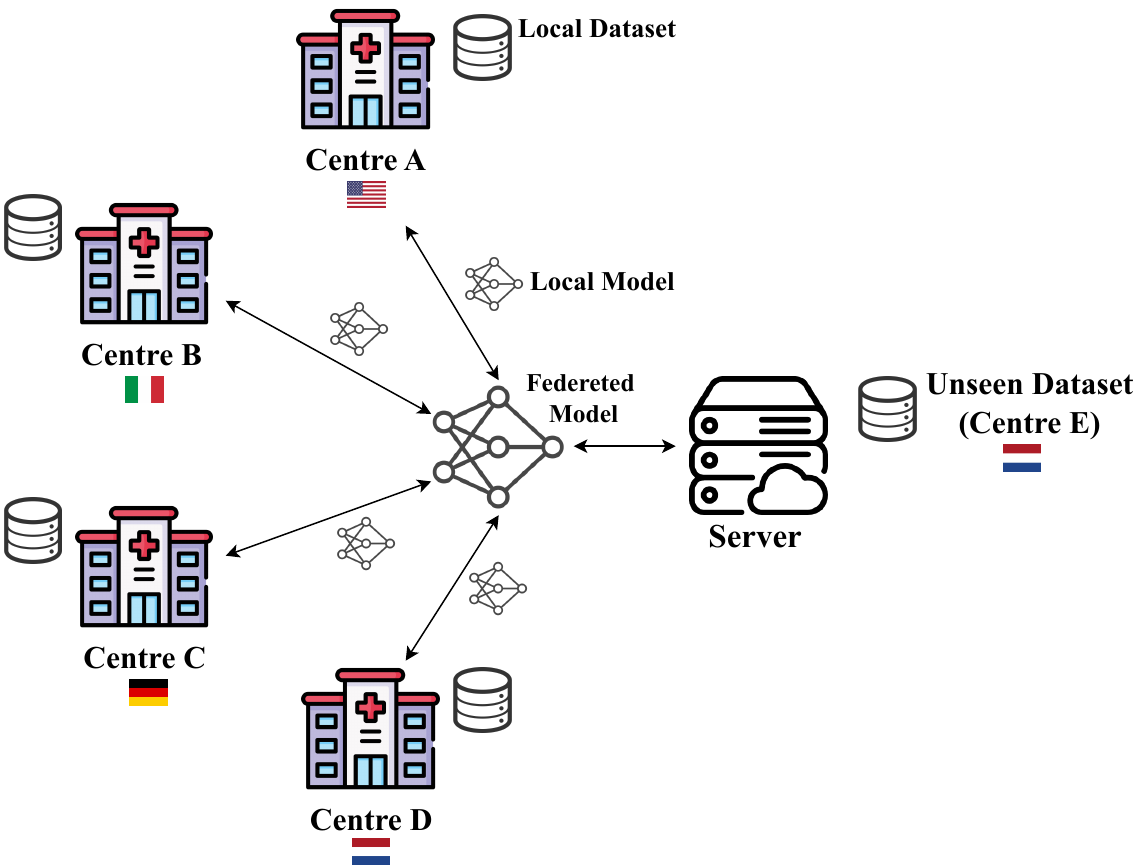}
    \caption{Federated setup. Each centre (also called client) has its own local data and trains a model on the local dataset. The weights of the local model are exchanged with the server, which, after the aggregation process, returns a federated model to the clients. The server also hosts a dataset that is not used during training, but acts as a benchmark to determine the generalisation ability of the model.}
    \label{fig:fl_setup}
\end{figure}

As shown in Figure \ref{fig:fl_setup}, we created a FL environment to simulate a realistic collaboration involving 4 distinct clients, each representing a different clinical centre as outlined. In this setup, each client trained a local DL model on its own data, while a central server coordinated the creation of a federated model. This federated model was built by aggregating key parameters (e.g., weights, biases, and batch normalization parameters) from each local model after every local training round, effectively combining weights from all clients without sharing raw data. The aggregated model parameters were then redistributed to each client, enhancing their local models with information learned across all sites.

To evaluate generalisation, the federated model was tested on an independent dataset from a fifth centre, which none of the clients had access to during training.

\subsection{Deep learning model}
\label{subsec:met_model}
\begin{figure}[tbp]
    \centering
    \includegraphics[width = 1\linewidth]{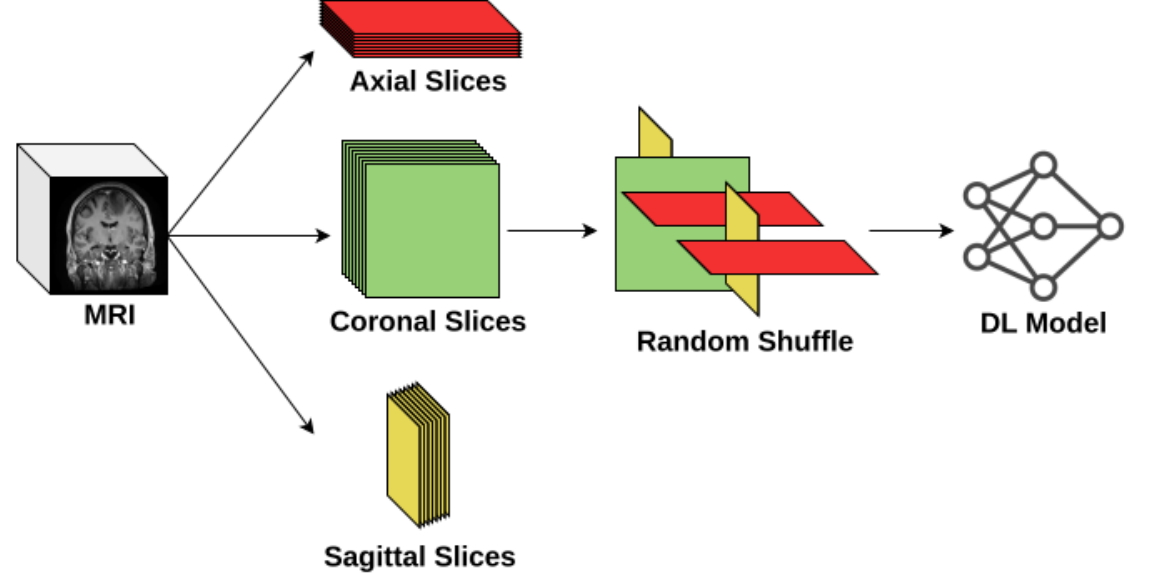}

    \caption{Proposed training methodology. The pre-processed and augmented 3D (MRI) volume is subdivided into 2D slices, extracted along the three anatomical planes (sagittal, axial, coronal). These slices are then subjected to a random shuffling process, ensuring that the model is agnostic with respect to both the anatomical plane and the imaging sequence. The shuffled batch of images is then provided as input to the model.}
    \label{fig:train_method}
\end{figure}

Following preliminary experiments (see Section \ref{sota_model_comparison}), the DL architecture employed in the proposed FL framework was the U-Net model proposed by Li et al. \cite{Li2019}. The network is a residual learning based U-Net, designed to predict $256\times256$ sCT 2D slices from $256\times256$ MR inputs. The architecture comprises 34 convolutional layers, starting with a $7\times7$ convolutional layer for initial feature extraction, followed by multiple $3\times3$ convolutional layers. 

A key feature of the architecture dealt with the inclusion of residual blocks, which consisted of two $3\times3$ convolutional layers, each followed by batch normalization and rectified linear unit (ReLU) activation, and a short skip connection that added the input of the block to its output to improve gradient propagation and enable efficient training. In the encoder path, max-pooling operations were used for downsampling the dimensions of the feature maps, doubling the number of filters at each step. In the decoder path, upsampling operations were performed before convolutional layers to restore feature maps to their original size, halving the number of filters at each step. Skip connections between the encoder and decoder paths facilitated the reuse of spatial information.

To train the aforementioned architecture, each client input 2D slices from each MRI in 3 anatomical planes, axial, coronal, and sagittal, as it is shown in Figure \ref{fig:train_method}. Slices were presented to the network in a randomised order. After extracting all slices from each anatomical plane, they were shuffled to ensure that the DL model received non-sequential input from more than one anatomical plane at a time. This minimised the risk of the model becoming overly dependent on any single plane or orientation, encouraging comprehensive and plane-invariant feature learning across all planes \cite{kamnitsas2017efficient}. This method was adapted from the Multi-2D approach \cite{Spadea2021}, modified to enhance generalisation. As a result, it was renamed Random Multi-2D.

\subsection{Voting}

\begin{figure}[tbp]
    \centering
    \includegraphics[width = 1\linewidth]{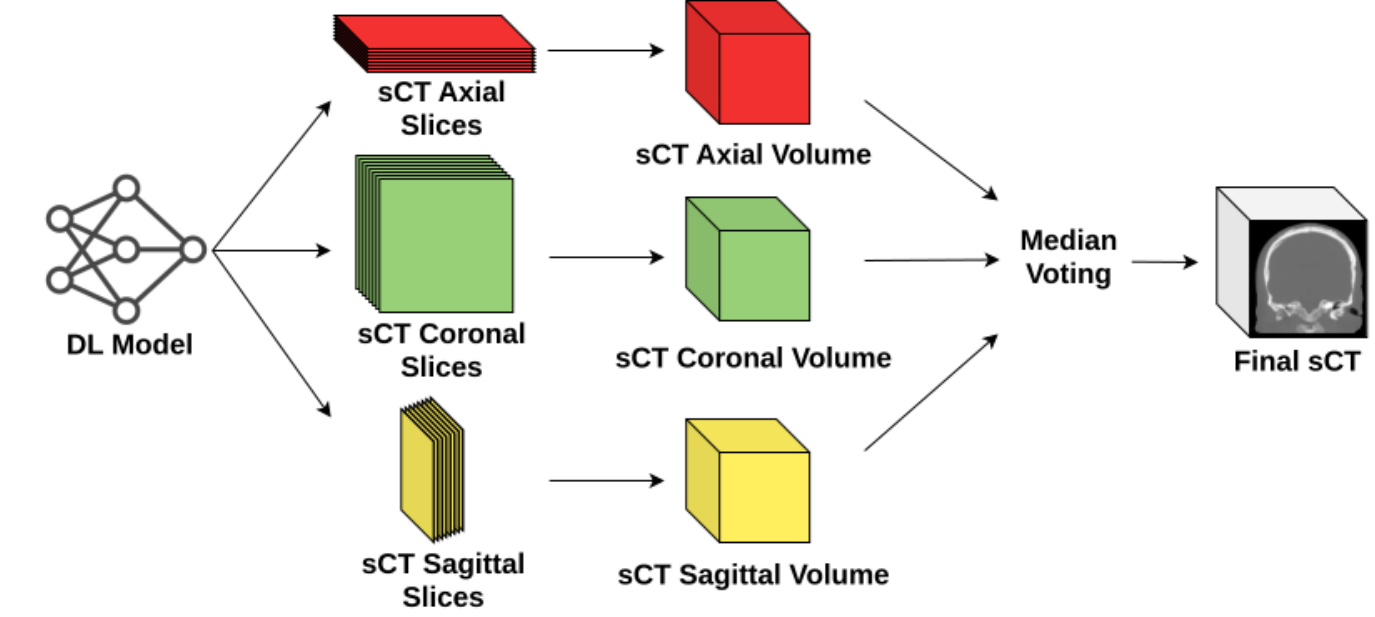}

    \caption{Reconstruction of the sCT. The axial, coronal, and sagittal slice predictions are rearranged to construct three distinct volumes. Subsequently, a median voting approach is adopted using the three volumes in order to enhance the final result.}
    \label{fig:voting_method}
\end{figure}

As shown in Figure \ref{fig:voting_method}, in order to reconstruct the sCT, the predictions generated by the three anatomical planes were combined using a median voting approach to obtain the final voxel value, as proposed in \cite{SpadeaDCNN2019}. This post-processing procedure ensured a more uniform and reliable output and was applied on both the server-side and the client-side predictions before the evaluation process. Specifically, once the predictions for each slice had been generated for each anatomical plane, the median result across the three axes was used to generate the final sCT voxel according to the formula~\cite{SpadeaDCNN2019}:
\begin{equation} \label{sCTvoting}
    sHU_{V_n} = \text{median}(sHU_{\text{Cor}_n}; sHU_{\text{Ax}_n}; sHU_{\text{Sag}_n}), \quad n \in \{1, \dots, N\}
\end{equation}

where $V_n$ (with $n \in \{1, \dots, N\}$) refers to the pixel number, $sHU$ refers to the synthetic Hounsfield Unit value of the $n$ pixel for the coronal, axial, and sagittal sCT.

\subsection{Aggregation strategy}

After testing different aggregation strategies (see section 4.4.2), FedAvg in combination with FedProx to address the challenges associated with non-IID data across clients \cite{li2020federatedepochsopt}, was adopted for FedSynthCT-Brain.  

More in detail, the weights from each client are aggregated round-by-round using a weighted average (thus, FedAvg), with each client's contribution proportional to the size of its dataset.

\begin{equation} \label{eq:fedavg}
w_{t+1} = \sum_{k=1}^{K} \frac{n_{k}}{n}w_{t+1}^k
\end{equation}

where \( K \) represents the number of clients, \( n_k \) denotes the dataset size for client \( k \), \( n \) is the total dataset size across all clients, and \( w_{t+1}^k \) are the weights of client \( k \) post-training. 

As FedAvg underperform with non-IID data \cite{li2019fedavgconvergence}, FedProx was incorporated to address this limitation. The FedProx method, specifically designed for heterogeneous federated environment, introduces a proximal term \( \frac{\mu}{2}||w-w_{t}||^2 \) to the local objective function, penalising large deviations from the global model weights to control local update variations:

\begin{equation} \label{eq:proximalterm}
h_{k}(w,w_{t}) = F_{k}(w) + \frac{\mu}{2}||w-w_{t}||^2
\end{equation}

where \( F_{k}(w) \) is the original local objective, \( w_t \) is the pre-training global model weight, \( w \) represents weights during local training, and \( \mu \) is the proximal coefficient, within the range $[0, 1]$. Therefore, FedProx reduces update divergence and approximates FedAvg when \( \mu = 0 \) \cite{li2020federatedepochsopt}.

\section{Experimental protocol}
\label{sec:expprot}

\subsection{Data preprocessing details}


The pre-processing pipeline used for each dataset can be divided into different steps, as shown in pseudo-code (Algorithm \ref{alg:preprocessing}). To correct the bias field, the N4ITK algorithm \cite{Tustison2010} was employed. The algorithm was applied to each MRI with identical parameters. 

\begin{algorithm}[H]
\caption{Pre-processing pipeline for each Centre in pseudo-code}
\label{alg:preprocessing}
\begin{algorithmic}[]
\If{Volume is MRI}
    \State Apply N4 Bias Field Correction
\EndIf
\If{Volume Size $>$ $256 \times 256 \times 256$}
    \State Apply Crop to centre the volume
    \State Resize to $256 \times 256 \times 256$ 
\EndIf
\If{Volume Size $<$ $256 \times 256 \times 256$}
    \State Pad to $256 \times 256 \times 256$ 
\EndIf
\If{Volume is MRI}
    \State Apply Min-Max Normalization
\EndIf
\end{algorithmic}
\end{algorithm}

The second step was the standardisation of the anatomical axes orientation, which facilitates the application of the subsequent transforms and the extraction of slices. Then, all volumes that had a larger dimension than the target dimension were centred cropped to a size of $328 \times 256 \times 328$. Finally, all volumes were reshaped to a voxel size of $1 mm \times 1 mm \times 1 mm$.

Furthermore, the resizing and padding process was executed by establishing a maximum dimension of $256$ for all directions. The padding values were $-1000$ HU and $0$ for CTs and MRIs, respectively.

Finally, the intensity values of each MRI were normalised to the $[0, 1]$ range using the Min-Max Normalization.

All experiments proposed in the subsequent sections were conducted with 4 clients, where each client used 2 patients for validation, 2 patients for testing, and the remaining for training.

\subsection{Implementation and training settings}

\label{sec:impl}

The implementation of the proposed work, was carried out in Python, using the Flower framework~\cite{beutel2022flowerfriendlyfederatedlearning} for the FL infrastructure and the PyTorch~\cite{pytorch} library for the DL implementation. To reproduce the federated environment and train the models, we used an NVIDIA A100 80GB GPU, 16 CPU cores, and 64GB of RAM from a Supermicro AS-4124GS-TNR 4U Rackmount GPU SuperServer. 

To maintain consistency and prevent overfitting, a custom-built data augmentation pipeline was implemented for each client. The AugmentedDataLoader library \footnote{\url{https://github.com/ciroraggio/AugmentedDataLoader}} was used to perform optimised on-the-fly data augmentation during training. In order to address the considerable computational demands associated with this data augmentation approach, two distinct data augmentation pipelines were employed: (i) a minimal pipeline, comprising a flip and a fixed-angle single-axis rotation; (ii) an extended pipeline, comprising a flip, a random-angle three-axis rotation and a fixed spatial translation; based on the amount of data available for the individual site. 

A preliminary study was conducted to establish the optimal number of local epochs each client, thereby ensuring that the global model converged efficiently, rather than focusing exclusively on local optimisation. This involved an investigation into the impact of varying the number of local clients training epochs on the overall performance of the global model.

As reported in \cite{li2020federatedepochsopt}, it is beneficial to minimise the size of local updates in the context of FL. With this view, experiments were conducted using $1, 2, 3$, and $5$ epochs per round. It has been experimentally observed that an increasing number of epochs per round resulted in a proportional increase in computation time, with a degradation in image similarity metrics for the federated model. Consequently, the optimal number of local epochs was experimentally set to $1$.

Each client model was trained using the Adam optimizer and the L1 loss. After an initial experimental analysis, the learning rate (LR) was set to $10^{-4}$. The batch size was set to 32.

Each experiment ran for 30 rounds. This setup allowed for a rigorous and fair evaluation of both the model's ability to generalise across rounds and its performance on unseen data.

\subsection{Performance metrics}\label{subsec:performance_metrics}

In order to evaluate the efficacy of the federated model both on the clients test datasets (centres A, B, C, and D), and unseen test dataset (centre E) standard image similarity metrics in the MRI to CT translation task were employed \cite{Spadea2021, Boulanger2021}. These included:
\begin{itemize}
    \item Mean Absolute Error (MAE):
\[
\text{MAE} = \frac{1}{n} \sum_{i=1}^{n} \left| CT_i - sCT_i \right|
\]
where n is the total number of voxels in the region of interest.
\end{itemize}
\begin{itemize}
    \item Structural Similarity Index Measure (SSIM) \cite{Spadea2021}:
\[\text{SSIM} = \frac{(2 \mu_{\text{sCT}} \mu_{\text{CT}} + C_1)(2 \sigma_{\text{sCT,CT}} + C_2)}{(\mu_{\text{sCT}}^2 + \mu_{\text{CT}}^2 + C_1)(\sigma_{\text{sCT}}^2 + \sigma_{\text{CT}}^2 + C_2)}
\]
\[
C_1 = (k_1 L)^2, \quad C_2 = (k_2 L)^2
\]
\[
\mu = \text{Mean}, \quad \sigma = \text{Variance/Covariance}
\]
\[
L = \text{Dynamic range}, \quad k_1 = 0.01, \quad k_2 = 0.03
\] 
\end{itemize}
\begin{itemize}
    \item Peak Signal-to-Noise Ratio (PSNR) \cite{Spadea2021}:
\[
\text{PSNR} = 10 \cdot \log \left( \frac{MAX_{CT}^2}{\text{MSE}} \right)
\]
\end{itemize}

We further evaluated the efficiency using the time elapsed to complete a fixed number of rounds. This was aimed at identifying the most suitable architecture for the FL framework, balancing generalisation capability and training time. Indeed, minimizing training time is crucial, as extended training may exacerbate resource constraints, increase communication overhead among clients, and compromise the scalability of the system, particularly in multi-institutional collaborations \cite{chen2024cost}.

\subsection{Deep learning architecture and aggregation strategy selection}
\subsubsection{Deep learning architecture selection}
\label{sota_model_comparison}

We explored different DL architectures used for MRI-to-sCT conversion, including Conditional GAN, CycleGAN and Transformers (see the supplementary \ref{app:arch_investigation} for further details). Among these, we further investigated several variants of the U-Net architecture which has been used for centralised MRI-to-sCT-Simple U-Net \cite{Ronneberger2015}, Spadea et al. U-Net \cite{SpadeaDCNN2019}, Fu et al. U-Net \cite{Fu2019}, Li et al. U-Nets \cite{Li2020}. These variants included variations in implementations, such as changes in network depth, feature map sizes, and the integration of attention mechanisms. Our analysis focused on evaluating the balance between computational complexity and generalization capability across these architectures. 

An additional investigation is presented in \ref{app:dl_methods_investigation} which involved a comparison of the adopted training methodology (Random Multi-2D) with other methodologies (such as 2D+, 2D patches and classic Multi-2D) using the U-Net presented in Section \ref{subsec:met_model}.

\subsubsection{Aggregation strategy selection}

Once the most appropriate architecture was identified, a comparative analysis was conducted between the chosen aggregation strategy (i.e., FedAvg+FedProx), simple aggregation methods, such as FedAvg, and optimised and more complex techniques, such as FedYogi and FedAvgM with the incorporation of the momentum coefficient and the server-side learning rate. 

Further optimisation was pursued through the incorporation of additional techniques, including FedBN \cite{li2021fedbn}, and the combination of FedProx and FedBN, with the aim of refining and optimising the federated training process. In our study, the FedBN method was specifically designed by the exclusion of batch normalization layers from communication between the server and client. This was done with a particular focus on ensuring that the global model could converge more rapidly and improve the robustness.

To determine the range of variability of the convergence trend within 25 rounds, the experiments were repeated 5 times for each strategy.

Regarding the hyper parameters used for each strategy, FedAvgM required the tuning of the momentum coefficient $\beta$ and the server-side learning rate $\eta_{s}$. It was noted experimentally that the use of $\beta > 0.4$ and $\eta_{s} < 0.1$ leads to instability due to the inability of the model to reach the convergence. In light of this, the parameters used for FedAvgM were $\beta = 0.3$ and $\eta_{s} = 0.2$, respectively.

The FedYogi strategy necessitates the configuration and optimisation of different hyper parameters. These have been empirically optimised following a series of experiments with the objective of stabilising convergence. The best result for FedYogi  was achieved with the following parameters: $\eta=0.03$, $\eta_l = 10^{-4}$, $\beta_1 = 0.6$,  $\beta_2 = 0.6$, and $\tau = 0.01$.

To add FedProx, different values of the proximal coefficient $\mu$ were tested. Higher $\mu$ corresponds to a greater penalty for significant deviations from the global model during local training, which may impede the model's convergence. On the other hand, setting $\mu = 0$ would be equivalent to FedAvg. The optimal results for FedProx were obtained with a $\mu = 3$.  

The FedBN approach was also tested with FedAvg and combined to FedAvg and FedProx. While no parameter tuning was required for FedBN, the conventional approach was modified. 

Batch normalisation layers were excluded solely when transmitting the global model to clients, thereby enabling the aggregation of all layers, including batch normalization, at the server site. This alteration guarantees that the global model can leverage client-specific batch normalisation updates, enhancing its resilience when evaluated with previously unseen data.

\section{Results}
\label{sec:res}

\subsection{Benchmarking deep learning architectures}
The performances of different state-of-the-art U-Nets, using a FL setup and the chosen aggregation strategy (FedAvg+FedProx), are presented in Table \ref{tab:unets_study_results}, in terms of median and inter-quartile range values of MAE, SSIM, and PSNR. The results are provided the test dataset of each centre participating to the federation (A, B, C, and D) and for the external centre E.

Image similarity assessment did not reveal any relevant difference across the tested U-Net architectures. However, when examining the total time required to complete 30 rounds, the U-Net architecture proposed by \cite{Li2019} required 12 hours and 34 minutes.

Additional results obtained with different architectures, which proved to be less suitable than U-Nets for the proposed federated task, are presented in Table \ref{tab:model_investigation}. Nevertheless, the results obtained from the supplementary investigation on the performance obtained of the U-Net by Li et al. \cite{Li2019} when using different DL training methodologies are reported in Table \ref{tab:train_paradigm_investigation}.

\begin{table}[tbp!]
    \centering
    \small
    \caption{Comparison of the results obtained with different state-of-the-art U-Nets on the unseen centre dataset and federated centres (A, B, C, D) test data based on image similarity metrics and training time, considered as the total time to complete a fixed number of rounds.}
    \label{tab:unets_study_results}
    \begin{tabular}{lcccccc}
        \toprule
        & \textbf{Method} & \textbf{MAE [HU]} & \textbf{SSIM} & \textbf{PSNR [dB]} & \textbf{Total Time} \\
                                 
        \midrule
        \multirow{5}{*}{\textbf{Centre A}}
        & Simple U-Net \cite{Ronneberger2015}      & 85.6 (84.8-86.4) & 0.96 (0.96-0.96) & 33.91 (33.65-34.18) & \multicolumn{2}{c}{-} \\
        &  Spadea et al.\cite{SpadeaDCNN2019}    & 88.6 (88.0-89.3) & 0.95 (0.95-0.95) & 33.72 (33.45-33.99) & \multicolumn{2}{c}{-} \\
        & Fu et al. \cite{Fu2019}        & 91.3 (90.7-92.0) & 0.95 (0.95-0.95) & 33.43 (33.15-33.71) & \multicolumn{2}{c}{-} \\
        & Li et al. (2020) \cite{Li_Comparison2020} & 93.0 (92.6-93.4) & 0.95 (0.95-0.95) & 33.34 (33.07-33.61) & \multicolumn{2}{c}{-}\\
        & Li et al. (2019) \cite{Li2019} & 89.7 (89.4-90.1) & 0.95 (0.95-0.95) & 33.64 (33.46-33.81) & \multicolumn{2}{c}{-} \\
        
        \midrule
        \multirow{5}{*}{\textbf{Centre B}} 
        & Simple U-Net \cite{Ronneberger2015}       & 126.1 (125.6-126.6) & 0.86 (0.86-0.86) & 21.83 (21.06-22.61) & \multicolumn{2}{c}{-} \\
        &  Spadea et al.\cite{SpadeaDCNN2019}    & 128.6 (128.1-129.0) & 0.86 (0.85-0.86) & 21.81 (21.04-22.59) & \multicolumn{2}{c}{-} \\
        & Fu et al. \cite{Fu2019}        & 125.7 (124.8-126.7) & 0.86 (0.86-0.86) & 21.82 (21.04-22.61) & \multicolumn{2}{c}{-} \\
        & Li et al. (2020) \cite{Li_Comparison2020} & 131.0 (130.5-131.5) & 0.86 (0.85-0.86) & 21.77 (21.01-22.54) & \multicolumn{2}{c}{-} \\
        & Li et al. (2019) \cite{Li2019} & 124.7 (124.4-125.1) & 0.86 (0.86-0.86) & 21.85 (21.08-22.63) & \multicolumn{2}{c}{-} \\

        \midrule
        \multirow{5}{*}{\textbf{Centre C}} 
        & Simple U-Net  \cite{Ronneberger2015}      & 97.0 (94.1-99.9) & 0.74 (0.71-0.77) & 20.03 (19.36 - 20.71) & \multicolumn{2}{c}{-} \\
        &  Spadea et al.\cite{SpadeaDCNN2019}    & 95.3 (92.0-98.6) & 0.74 (0.71-0.77) & 20.06 (19.37 - 20.73) & \multicolumn{2}{c}{-} \\
        & Fu et al. \cite{Fu2019}        & 89.1 (83.8-94.3) & 0.74 (0.72-0.77) & 20.05 (19.38 - 20.75) & \multicolumn{2}{c}{-} \\ 
        & Li et al. (2020) \cite{Li_Comparison2020} & 97.1 (95.6-98.6) & 0.74 (0.72-0.77) & 20.03 (19.36 - 20.70) & \multicolumn{2}{c}{-} \\ 
        & Li et al. (2019) \cite{Li2019} & 92.9 (89.8-96.0) & 0.74 (0.72-0.77) & 20.05 (19.37 - 20.73) & \multicolumn{2}{c}{-} \\
        
        \midrule
        \multirow{5}{*}{\textbf{Centre D}} 
        & Simple U-Net \cite{Ronneberger2015}       & 70.8 (66.9-74.7) & 0.93 (0.93-0.93) & 27.05 (27.03-27.08) & \multicolumn{2}{c}{-} \\ 
        &  Spadea et al.\cite{SpadeaDCNN2019}    & 71.7 (68.8-74.7) & 0.93 (0.93-0.93) & 27.05 (27.03-27.06) & \multicolumn{2}{c}{-} \\ 
        & Fu et al. \cite{Fu2019}        & 71.7 (68.4-75.0) & 0.93 (0.93-0.93) & 27.03 (27.01-27.05) & \multicolumn{2}{c}{-} \\ 
        & Li et al. (2020) \cite{Li_Comparison2020} & 71.6 (66.5-76.7) & 0.93 (0.93-0.93) & 27.04 (27.00-27.07) & \multicolumn{2}{c}{-} \\ 
        & Li et al. (2019) \cite{Li2019} & 72.6 (63.4-76.8) & 0.93 (0.93-0.93) & 27.04 (27.02-27.06) & \multicolumn{2}{c}{-} \\ 

        \midrule
        \multirow{5}{*}{\makecell{\textbf{Centre E} \\ \textbf{(Unseen)}}} 
        & Simple U-Net   \cite{Ronneberger2015}     & 100.3 (94.5-108.2) & 0.89 (0.86-0.89) & 26.56 (25.53-27.52) & 16h 22min \\
        & Spadea et al.\cite{SpadeaDCNN2019}    &  99.8 (92.6-108.9) & 0.89 (0.86-0.89) & 26.56 (25.53-27.52) & 22h 41min \\
        & Fu et al. \cite{Fu2019}        &  99.8 (94.9-106.1) & 0.89 (0.86-0.89) & 26.61 (25.53-27.49) & 20h 22min \\
        & Li et al. (2020) \cite{Li_Comparison2020} &  98.9 (93.3-108.0) & 0.89 (0.86-0.89) & 26.55 (25.57-27.45) & 15h 57min \\
        & \textbf{Li et al. (2019)} \cite{Li2019} & 102.0 (96.7-110.5) & 0.89 (0.86-0.89) & 26.58 (25.52-27.42) & \textbf{12h 34min } \\
        
        \bottomrule
    \end{tabular}
    
\end{table}

\subsection{Benchmarking aggregation strategies}
Table \ref{tab:strategies_comparison} presents the best result obtained for each aggregation strategy, tested on the unseen dataset (centre E). Results are reported  in terms of MAE, which resulted the most sensitive image similarity metric in the benchmarking of DL architectures. The experiments (Experiment 1, 2, 3, 4, 5), were repeated 5 times for each strategy to determine the range of variability of the convergence trend within 25 rounds. This comparison helps to understand which aggregation strategy gives the best result in a stable and efficient way.  For each experiment, the number of rounds performed to obtain the best result is also reported. Figure \ref{fig:AggregationsPlots} reports the convergence plots for all evaluated strategies. Furthermore, a visual comparison of the results obtained with different strategies were reported in Figure \ref{fig:AggregationsVisualResults}.

\begin{table}[tbp!]
\centering
\small
\caption{Comparison, in terms of image similarity (MAE) and number of rounds to achieve the best results among different aggregation strategies. Experiments were repeated 5 times and the mean and standard deviation (Std) are presented for both number of rounds and the MAE.}
\begin{tabular}{lccccccc}
\toprule
\textbf{Strategy} & \textbf{Metric} & \multicolumn{5}{c}{\textbf{Experiment}} & \textbf{Mean$\pm $Std} \\
\cmidrule{3-7}
& & \textbf{1} & \textbf{2} & \textbf{3} & \textbf{4} & \textbf{5} & \\

\midrule
\multirow{2}{*}{FedAvg} & \makecell{Round} & 10 & 16 & 20 & 10 & 18 & $14.8 \pm 4.1$ \\ 
    & \makecell{MAE [HU]} & 99.5 & 99.3 & 100.8 & 102.6 & 101.1 & $100.7 \pm 1.4$ \\ 

\midrule
\multirow{2}{*}{FedAvgM} & \makecell{Round} & 28 & 27 & 28 & 29 & 27 & $27.8 \pm 0.7$ \\ 
    & \makecell{MAE [HU]} & 102 & 101.8 & 103 & 101 & 101.4 & $101.8 \pm 0.7$ \\ 

\midrule
\multirow{2}{*}{FedYogi} & \makecell{Round} & 13 & 13 & 17 & 4 & 12 & $11.8 \pm 4.3$ \\ 
& \makecell{MAE [HU]} & 207.7 & 230.6 & 225.9 & 236.1 & 234.4 & $226.9 \pm 10.2$ \\

\midrule
\multirow{2}{*}{\makecell{FedAvg +\\ FedBN}} &   \makecell{Round} & 20 & 6 & 15 & 21 & 21 & $16.6 \pm 5.7$ \\ 
& \makecell{MAE [HU]} & 101 & 100.9 & 99.6 & 100.7 & 98.3 & $100.1 \pm 1.1$ \\ 

\midrule
\multirow{2}{*}{\makecell{FedAvg + \\FedProx + \\ FedBN}} & \makecell{Round} & 15 & 8 & 15 & 21 & 19 & $15.6 \pm 4.4$ \\ 
& \makecell{MAE [HU]} & 99.2 & 99.9 & 97.5 & 100.1 & 100.1 & $99.36 \pm 1$ \\ 
\\

\midrule
\multirow{2}{*}{\makecell{\textbf{FedAvg} \textbf{+}\\ \textbf{FedProx } \\ \textbf{(Our)}}} & \makecell{Round} & 18 & 8 & 14 & 14 & 9 & $\textbf{12.6} \pm \textbf{3.7}$ \\ 
& \makecell{MAE [HU]} & 99.8 & 99.2 & 100.1 & 101.6 & 98.3 & $99.8 \pm 1.1$ \\ \\
\bottomrule
\end{tabular}

\label{tab:strategies_comparison}
\end{table}

\subsection{Federated sCTs}
\begin{figure}[tbp]
    \centering
    \includegraphics[width = 1\linewidth]{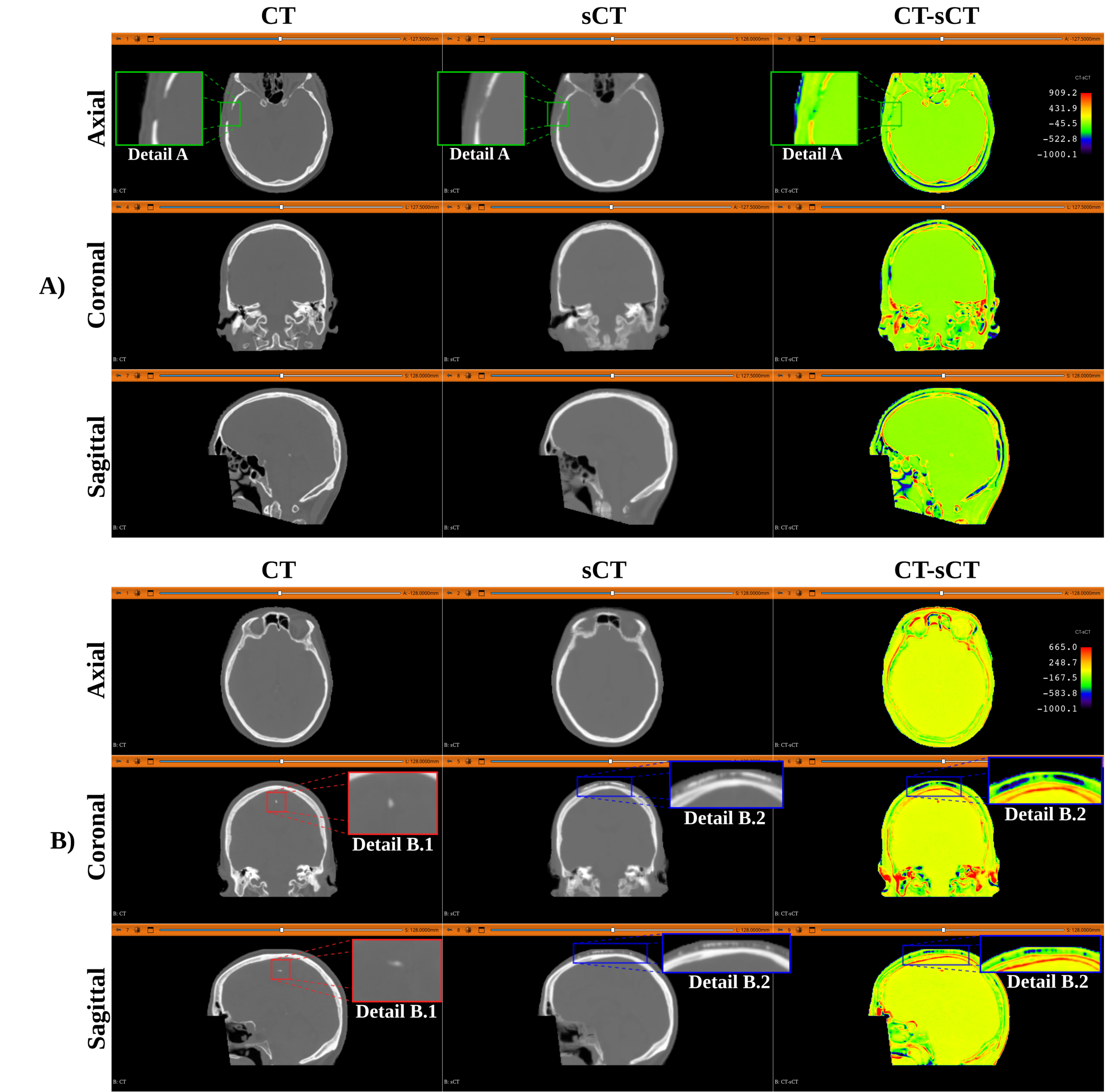}
    \caption{The figure shows the central slice for the axial, coronal and sagittal planes, which is used for an evaluation of the worst and best --A and B respectively-- predictions based on image similarity metrics of the federated model on the unseen dataset (Centre E). The Detail A (in green), shows the ability of the model to predict a missing portion of bone, although the case A was from the unseen Centre. In the case B, the Detail B.1 (in red) highlights a calcification in the ground truth CT, which is not correctly predicted by the model. In the case B, the Detail B.2 (in blue) also highlights a thin duplication of the upper skull bone. This issue has been identified exclusively within the predictions generated by the federated model on the Centre E dataset. The colour map on the left illustrates the differences between the CT and the sCT, highlighting regions where the model exhibits a higher or lower tendency to over or underestimate the actual outcome.}
    \label{fig:sCTsCentreE}
\end{figure}

\begin{figure}[tbp]
    \centering
    \includegraphics[width = 1\linewidth]{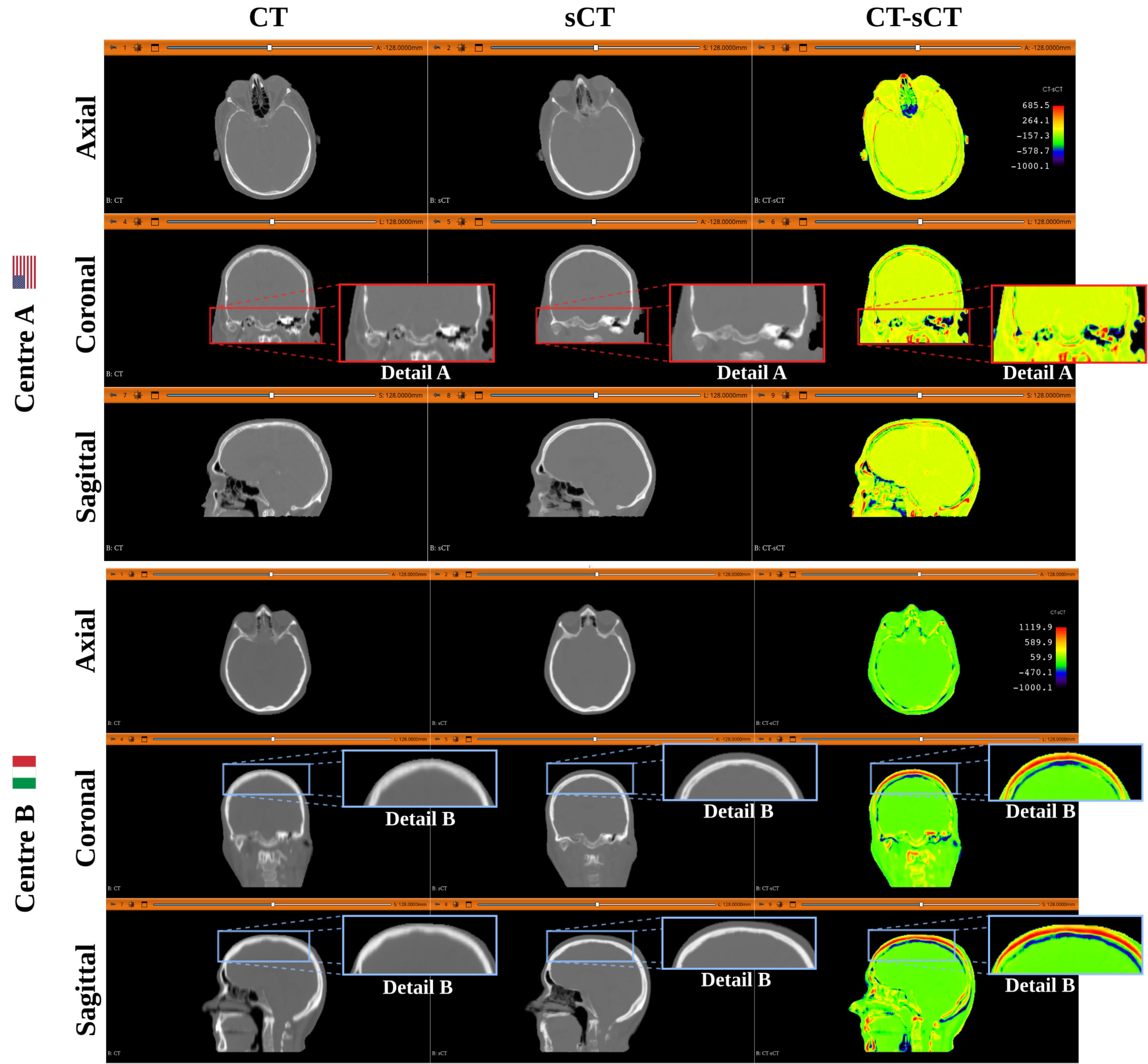}
    \caption{The figure shows the central slice for the axial, coronal, and sagittal planes, which is used for an assessment of one of the two test cases of Centres A and B, respectively. The Detail A (in red) comparison highlight the approximation of the small structure performed by the model during the sCT generation for Centre A. Detail B comparison (in blue) shows the presence of artefact in Centre B's ground-truth CTs, whereby the upper skull is larger than expected.
    Despite this artefact, the federated model predicts a smaller skull structure, and thus a more realistic outcome.}
    \label{fig:sCTsCentreAB}
\end{figure}

\begin{figure}[tbp]
    \centering
    \includegraphics[width = 1\linewidth]{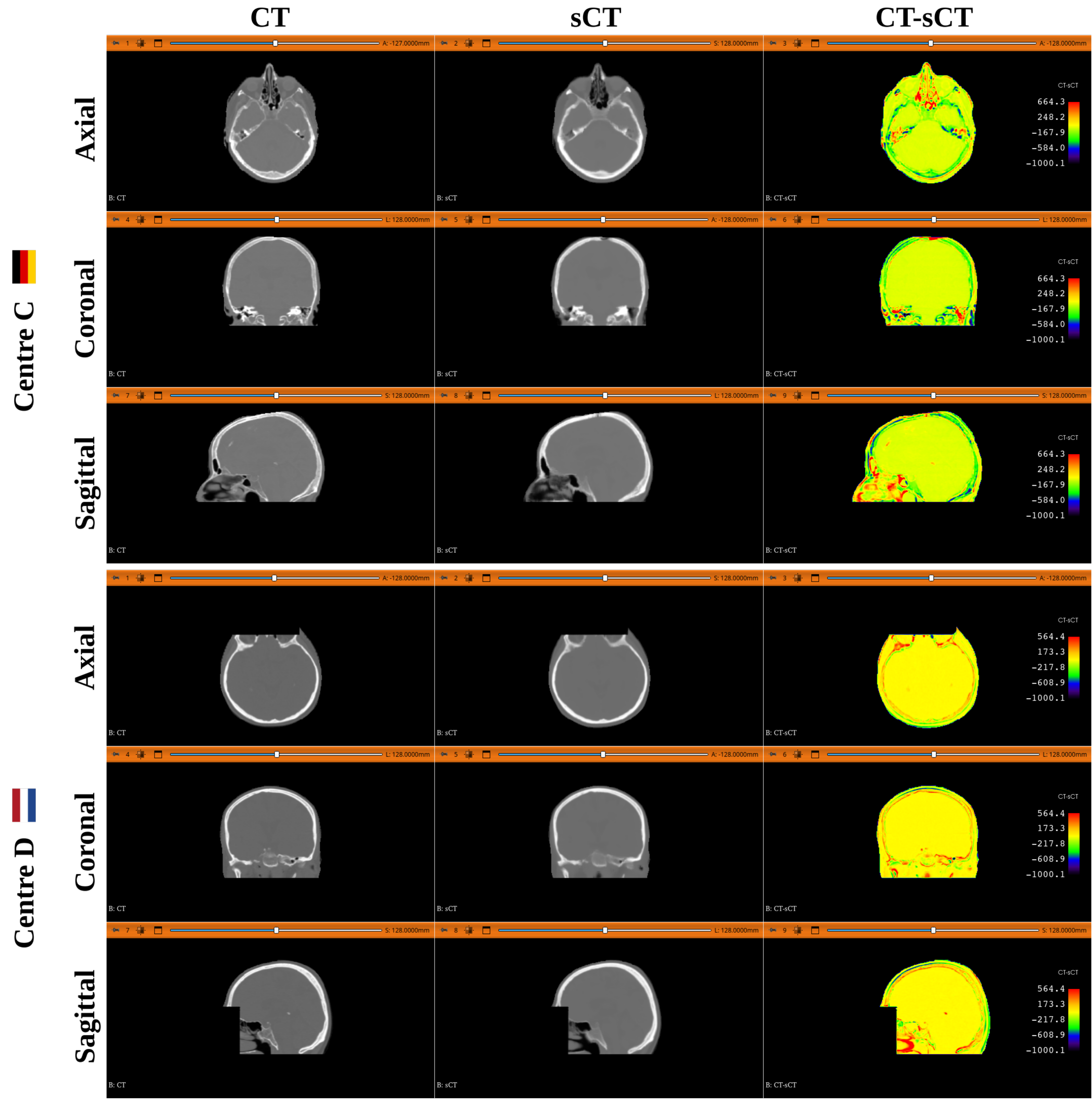}
    \caption{The results obtained from the federated model for one of the two test cases of Centres C and D. As for the other Centres of the federation, conversion problems emerge in small and detailed areas, while for the majority of the image composed of soft tissue the error remains contained.}
    \label{fig:sCTsCentreCD}
\end{figure}

As reported in Table \ref{tab:unets_study_results}, using the U-Net by Li et al. \cite{Li2019} and FedAvg combined with FedProx as aggregation strategy, the median MAE (inter-quartile range) for the unseen dataset was $102.0$ HU ($96.7-110.5$ HU).

Figure \ref{fig:sCTsCentreE} shows results for the best and worst cases of Centre E.

A view of the central slice in the three anatomical places is provided, demonstrating the advantages and limitations of the framework. Furthermore, the difference between the ground truth CT and the sCT is shown to assess critical pattern in the sCT generation. 

The results obtained by the federated model on one random patient of each client test dataset are also presented for each centre in Figures \ref{fig:sCTsCentreAB} and \ref{fig:sCTsCentreCD}. The central slice of the three anatomical planes is provided for the clients' data, along with the ground truth CT and sCT difference. The figures include rectangles of different colours to identify errors or artefacts observed in the images during the comparative analysis, that are not evident when considering only the similarity metrics.
\section{Discussion}
\label{sec:disc}

This study addressed the challenges of bias and limited diversity in medical imaging datasets caused by single-institution reliance in the context of MRI-to-sCT. Our proposal was to
introduce a FL framework designed for MRI-to-CT synthesis in brain imaging, FedSyntCT-Brain. We reproduced a realistic FL environment, adhering to all FL principles, using a cross-silo horizontal FL paradigm to enable the collaborative training of DL models as if multiple institutions were involved. To further validate the effectiveness of our approach, we tested the performance of the federated model on an unseen dataset, demonstrating its generalisability to external heterogeneous data sources.

Regarding the evaluation of U-Net architectures, the results obtained on Centre E — the unseen dataset — as well as across the various Centres in the federation, reveal no relevant differences in image similarity metrics when comparing different architectures. Based on the MAE metric, the minimal variation in HU does not allow for the identification of a clearly superior U-Net design. This could imply that, within the context of the FedSynthCT-Brain framework, the architectures are sufficiently robust to handle the data variability across the different centres. Furthermore, SSIM and PSNR, being less sensitive to small variations than MAE, remained stable when comparing different architectures within the same centre. Consequently, the focus shifted to evaluating the disparity in efficiency among the different implementations, as this becomes a critical factor in determining the most suitable model for FL applications. 

Referring to results about the time taken to complete the experiment for a fixed number of rounds (Total Time, see Table \ref{tab:unets_study_results}), the U-Net proposed by Li et al. \cite{Li2019} was the most efficient. These findings are relevant in federated environments, where limited computational resources and the time spent on computation become major factors. Consequently, a model that requires less time to be trained and federated, might mitigate interruptions in the communication during training and minimise synchronisation times. 

The employment of U-Net architectures was further justified by supplementary experiments comparing various models, including Conditional GAN, CycleGAN and Transformers. The results of these experiments, which are discussed in \ref{app:arch_investigation}, demonstrated that the U-Net achieved the optimal overall performance, balancing image similarity metrics and computational efficiency, thus rendering them more suitable for FL MRI-to-sCT applications. \newline

Moving to the aggregation strategies, the analysis of FedAvg results (see Table \ref{tab:strategies_comparison}, revealed that the optimal performance were obtained in 25 rounds, exhibiting minimal variation -- $1.4$ HU -- in terms of MAE. However, when compared to the proposed aggregation method (FedAvg+FedProx), FedAvg required an average higher number of rounds, indicating a longer time to reach its optimal result. Referring to the Table \ref{tab:strategies_comparison}, using FedProx enhances the performance of FedAvg and accelerates the convergence process, achieving its optimal result in an average of $12.6 \pm 3.7$ rounds, compared to $14.8 \pm 4.1$ rounds for the FedAvg alone. This finding aligns with previous studies in closer fields of research, which highlight that the introduction of a proximal term can mitigate the impact of client heterogeneity and improve convergence stability in FL settings.

Despite the introduction of the momentum coefficient and the server-side learning rate, the FedAvgM strategy was unable to improve the performance in terms of image similarity. Also in this case, a minimal stochastic variation in terms of MAE was observed. As illustrated
in both Figure \ref{fig:AggregationsPlots} and Table \ref{tab:strategies_comparison}, while the trend of convergence remains consistent across experiments, FedAvgM requires $27.8 \pm 0.7$ (mean $\pm$ std) rounds, thereby doubling the number of rounds to achieve the optimal result in comparison to the proposed aggregation strategy. Consequently, it can be deduced that the momentum coefficient and the server-side learning rate do not offer any advantages in terms of performance or image quality when compared to the proposed method. Similar challenges with FedAvgM have been reported in \cite{reddi2020adaptive}, where the additional hyperparameters introduced by momentum strategies were found to complicate tuning without consistently improving performance.

The more complex and optimised strategy FedYogi, presented several stability problems in the performed experiments. Despite the efforts made to tune the hyper-parameters of the strategy, it did not yield comparable or acceptable results compared to simpler strategies, as evidenced by the average MAE of $226.9 \pm 10.2$.

From Table \ref{tab:strategies_comparison}, a delay in the convergence of the simpler FedAvg has been noted when introducing FedBN, in contrast to the approach of combining only FedProx. 

To confirm the benefit of the proximal term in this specific federated task, a combination of FedProx with FedAvg and FedBN (FedAvg + FedProx + FedBN) was investigated. It is evident from Table \ref{tab:strategies_comparison} that, despite the lack of improvement, the introduction of the proximal term (FedProx) resulted in a slight reduction in the average number of rounds required for convergence. Consequently, we concluded that the chosen strategy, which involves the simple combination of FedAvg with FedProx, with a proximal coefficient $\mu = 3$, offers a comparable performance to more complex and optimised strategies, while requiring fewer rounds. 

As in the case of the U-Nets evaluation, our findings reaffirm previous insights into the balance between model efficiency and efficacy \cite{li2020federatedepochsopt} while highlighting opportunities for future exploration, particularly in aggregation strategies that improve robustness and resource efficiency in FL scenarios.

Regarding the generalisation ability of the federated model, the evaluation of CT and sCT similarity on the unseen dataset (see  Figure \ref{fig:sCTsCentreE}) revealed that for the predominant portion of the image, specifically the soft tissue, the discrepancy ranges between -100 HU and 100 HU, with peak errors occurring in specific and limited regions, such as the nasal and oral cavities.

Some inconsistencies were identified in the upper skull of a few patients, indicated by the Detail B.2 in Figure \ref{fig:sCTsCentreE}, case B). The wrong I2I translation produced a subtle bone mirroring. In other cases the federated model was not able to predict the presence of calcifications in the brain tissue. This failure can be attributed to the limited number of examples that exhibited this brain calcification in the federation, which precluded the model from identifying abnormal cases such as the one highlighted. 

Despite the Centre E's dataset being external to the training, the federated model demonstrated its capacity to predict a missing portion of bone in the axial view (see Figure \ref{fig:sCTsCentreE}, case A, Detail A).

The results of the individual centres presented in Figure \ref{fig:sCTsCentreAB} and Figure \ref{fig:sCTsCentreCD}, exhibited various advantages and disadvantages, given the differing characteristics of the data from each Centre.

In Figure~\ref{fig:sCTsCentreAB} the aforementioned issue identified in the upper part of the skull for some cases of the Centre E does not occur.
However, the smaller and more detailed areas tended to be approximated, resulting in a loss of detail, particularly in the proximity of the nasal cavities, as can be note within the red box provided in the coronal view for the test case of the centre A.

As described in Section~\ref{subsec:data_preparation}, Centre B
was introduced in the federation to increase data heterogeneity, as the MRIs were acquired with a hybrid PET-MRI scanner, with larger CT slice thickness and different X-ray tube energy (120 kV vs. 140 kV). From Table~\ref{tab:unets_study_results}, it is evident that the error for the Centre B is greater than for the other clients, emphasising the difference from the other images in the federation. However, it can be observed that the error is partly due to the lower spatial resolution of the ground truth CT, introducing partial volume artefacts in the upper part of the skull (see Figure~\ref{fig:sCTsCentreAB}, Detail B). The result of this artefact is a noticeable, unrealistict thickening of the bone, as presented in Figure \ref{fig:CentreBArtefact}. Despite this problem, the skull bone of the predicted federated sCT appears more realistic and consistent with other examples. 
This establishes an advantage of the federation, as the influence given by the weights of the other Centres allows the federated model to remain robust in front of optimization errors during training. 

In Figure~\ref{fig:sCTsCentreCD}, results obtained for Centres C and D test cases are presented. The limitations of the model in reconstructing small, detailed parts are also evident in these cases. However, from the difference shown on the left, it can be observed that the error for the largest part of the images is always in the [-100,100] HU range, while the highest errors are consistently present in the proximity of the air and detailed zones. \newline

Furthermore, in order to support the validity of the proposed training strategy, the results obtained through alternative well-established training methodologies using the Li et al. model were presented and discussed in detail in \ref{app:dl_methods_investigation}, demonstrating the effectiveness of the chosen Random Multi-2D training approach.\newline

This work established the foundations for future developments. Several limitations of the employed architecture could be addressed by adopting more advanced medical image synthesis methodologies \cite{atli2024i2imambamultimodalmedicalimage, arslan2024selfconsistentrecursivediffusionbridge, kabas2024physicsdrivenautoregressivestatespace}, optimising their efficiency, and ensuring compatibility with the FL environment. A possible improvement could also involve optimizing the implementation of 3D architectures for federated scenarios by leveraging more efficient training strategies, such as model pruning, knowledge distillation, or dimensionality reduction \cite{soni2024federated}. Additionally, exploring hybrid federated approaches, such as cross-silo horizontal FL with asynchronous updates or client clustering \cite{nguyen2022federated, wang2022accelerating}, could reduce synchronization times. 

A further limitation of the present study was the available computational resources, which imposed a limit on the number of clients that could be incorporated into the federation. In future work, a real-world deployment could be considered to mitigate the constraints imposed by a single, although powerful, machine. In addition, the evaluation of the generated sCTs using the emerging sCT quality control algorithms \cite{Zaffino2024} has the potential to enhance the comprehension of the limitations of the federated framework and to introduce enhancements in future studies. 

Moreover, the implementation of specific pre-processing or data harmonisation techniques can be considered to enhance the homogeneity of the data and to mitigate potential hidden artefacts \cite{Han2025}. This, consequently, will improve the quality of the sCTs for both the Centres within the federation and in the generalisation task.

\section{Conclusion}
\label{sec:conc}
We proposed, for the first time to best of our knowledge, a FL approach for MRI to sCT, with a specific focus on the brain region. Despite the intrinsic complexity of the task, which was evaluated including real clinical datasets. The results illustrate the potential of this methodology to safeguard patient confidentiality while sustaining robust model performance. The incorporation of additional clinics and institutions into the federation will enhance the accessibility of diverse and heterogeneous data, thereby optimising the model's generalisability and performance in a secure and scalable manner. This approach allows for the creation of a model with the capacity to generalise to previously unseen datasets, which is a crucial aspect for wider clinical applicability. Furthermore, it underscores the viability of federated learning as a sustainable solution for medical imaging tasks.

Apart from the advantages that have been highlighted, the objective of this work was to establish the foundations for future research. Multiple approaches can be followed to enhance the final outcome and progressively reduce the necessity for data centralisation and the training of individual models for each clinic. Future developments will include dosimetric studies to establish whether the results obtained can be used also in the clinical environment for for external beam radiotherapy and targeted radionuclide treatments, as well as the benchmarking of aggregation methods that extend beyond classical methodologies but necessitate additional effort for implementation, such as FedFisher \cite{jhunjhunwala2024fedfisherleveragingfisherinformation} and FedDG \cite{Liu_Quande2021}. Furthermore, efforts will be made to integrate and optimise recent sCT methods employing complex 3D architectures.

\section*{Data availability}
The data used for the study acquired from centre A, B and C are confidential. The data from Centre D and Centre E were extracted from the public SynthRAD2023 Grand Challenge dataset and are available at \url{https://synthrad2023.grand-challenge.org/}.

\section*{Acknowledgments}
We would like to acknowledge Dr. C. Catana (Massachusetts General Hospital) for contributing to data sharing and support during this research.

\newpage
\appendix
\section{Investigation of deep learning architectures}
\label{app:arch_investigation}
This preliminary investigation was conducted using data from centres A and C, with Centre E used as the unseen dataset. This choice reduced computational time and allowed evaluation of multiple architectures. The hyper-parameters used are detailed in Section \ref{sec:impl}. The minimal augmentation pipeline was used to further reduce the computational time.
Among the selected architectures, were: U-Net (described in Section \ref{subsec:met_model}), Conditional GAN (Pix2Pix GAN, with a U-Net-based generator) \cite{isola2018imagetoimagetranslationconditionaladversarial}, and CycleGAN (with a ResNet-based generator) \cite{Zhu2017CycleGAN}, which are commonly used in image-to-image translation or CT synthesis tasks \cite{DAYARATHNA2024, Spadea2021, Boulanger2021}. Despite its inherent complexity, we also evaluated the winning deep learning (DL) model of the SynthRAD challenge, which used a patch-based approach for centralized MRI-to-sCT \cite{huijben2024generating}, namely the MSEP architecture \cite{chen2023hybrid}.\newline

\textbf{Pix2Pix Conditional GAN.}
The Pix2Pix conditional GAN (cGAN) model employed a U-Net-based generator and a PatchGAN discriminator, used to have a $30\times30$ probability patch in output rather than a single value. For this model, an hybrid loss that integrated the pixel-wise error of the prediction and the discriminator's assessment was employed. \newline

\textbf{CycleGAN.}
The CycleGAN used in this study was composed of two generators and two discriminators. The generator networks were implemented as instances of a ResNet, while PatchGANs with an output size of $30\times30$ were employed for the discriminators. The generator loss was calculated as a weighted average of three components: paired pixel-wise loss, cycle-consistency loss, and adversarial loss. \newline

\textbf{MSEP.}
To enable its use in a FL context, we adapted the original implementation accordingly. To train the MSEP, patches with dimensions $48\times160\times160$ were extracted from each sample. The location of each patch was chosen randomly every epoch to increase the amount of possible training input data needed by intrinsic nature of the architecture. Since with this 3D DL approach, the model predicted overlapping patches and the number of predictions was not constant across the volume, the average value was calculated for each voxel using the median voting approach to reach a consensus.

In this scenario, the hardware needed to lead the experiment was an NVIDIA A100 80GB GPU, 32 CPU cores, and 256GB of RAM from a Supermicro AS-4124GS-TNR 4U Rackmount GPU SuperServer.

\begin{table}[h]
    \centering
    \scriptsize
    \begin{tabular}{c c c c c}
        \toprule
        & \multicolumn{3}{c}{\textbf{Center E (Unseen)}} \\
         \cmidrule{2-5}
        \textbf{Model} & \textbf{MAE [HU]} & \textbf{SSIM} & \textbf{PSNR [dB]} & \textbf{Time} \\
        \midrule
        \textbf{U-Net} & \textbf{110.4 (91.2-126.1)} & \textbf{0.88 (0.85-0.89)} & \textbf{26.44 (25.42-27.27)} & 6h 00min \\
        Pix2Pix & 122.0 (104.6-139.1) & 0.87 (0.85-0.87) & 26.31 (25.27-26.98) & \textbf{4h 00min} \\ 
        CycleGAN & 121.4 (115.7-130.9) & 0.87 (0.85-0.88) & 26.30 (25.32-27.07) & 10h 28min \\ 
        MSEP & 240.5 (220.7-250.7) & 0.79 (0.77-0.80) & 23.70 (23.09-24.29) & 85h 37min\textbf{*}\\
        \bottomrule
    \end{tabular}
    \caption{The median and interquartile range (IR) for the mean absolute error (MAE), the structural similarity index (SSIM), and the peak signal-to-noise ratio (PSNR) obtained using different architectures, as well as the total time required for the experiments.\\{(\textbf{*}}) Time required to perform a single round.}
    \label{tab:model_investigation}
\end{table}

The results reported in \ref{tab:model_investigation} demonstrated that the U-Net obtained the optimal overall performance, taking into account both image similarity metrics and computational efficiency, with a training duration of 6 hours employing only Center A and C. 

Although the timing of Pix2Pix GAN was favourable, its conversion performance was inferior to that of U-Net under the same conditions. This difference was especially evident from the median and interquartile range of the MAE presented in Table \ref{tab:model_investigation}.

The CycleGAN model demonstrated inferior performance in terms of image similarity when employing equivalent settings, with a training duration of 10 hours and 28 minutes that was nevertheless less efficient than the U-Net. 

Regarding the MSEP transformer-based architecture, despite efforts to reduce computational overhead — such as limiting the number of participating clients and reducing the size of the 3D patches — the time required for a single federated round remained prohibitively high. Specifically, one round took approximately 85 hours and 37 minutes. Moreover, the federated MSEP architecture’s performance fell short when compared to its centralised counterpart. This decline can be primarily attributed to: (i) the adjustments made to the training pipeline to accommodate the federated framework, and (ii) the reduced number of rounds and clients necessitated by the computational complexity of the 3D approach.

\section{Training methods investigation}
\label{app:dl_methods_investigation}
The primary objective of this investigation was to further evaluate the effectiveness of the proposed FL training approach compared to well-established training methods described in the literature for the MRI-to-sCT task in a centralised setting, described in Section \ref{sec:centralised_mri_sct}. This investigation was applied to the selected model, U-Net by Li et al. \cite{Li2019}, which had previously demonstrated better performance in terms of efficiency when compared with different U-Nets. The hyper-parameters used are detailed in Section \ref{sec:impl}. The model was trained across Centres A, B, C, D, with Centre E serving as an unseen test dataset. The performance of each training methodology was assessed using the metrics described in Section \ref{subsec:performance_metrics}, computed on the unseen dataset. In the patch-based method, we extracted randomly cropped patches of $160\times160$ pixels to maximise contextual information. Overlapping regions were merged using pixel-wise averaging.

\begin{table}[h]
    \centering
    \scriptsize
    \begin{tabular}{c c c c c}
        \toprule
        & \multicolumn{3}{c}{\textbf{Center E (Unseen)}} \\
         \cmidrule{2-5}
        \textbf{Method} & \textbf{MAE [HU]} & \textbf{SSIM} & \textbf{PSNR [dB]} & \textbf{Time} \\
        \midrule
        2D+ & 104.0 (99.5-115.6) & 0.89 (0.86-0.89) & 26.49 (25.43-27.35) & 16h 00min \\ 
        2D Patches & 162.4 (155.3-173.0) & 0.84 (0.82-0.85) & 25.52 (24.65-26.12) & \textbf{8h 39min} \\ 
        Multi-2D & 111.7 (107.7-122.3) & 0.88 (0.85-0.89) & 26.33 (25.32-27.20) & 9h 35min \\
        \midrule
        \textbf{Random}\\\textbf{Multi-2D (Our)} & \textbf{102.0 (96.7-110.5)} & \textbf{0.89 (0.86-0.89)} & \textbf{26.58 (25.52-27.42)} & 12h 34min \\
        \bottomrule
    \end{tabular}
    \caption{The median and interquartile range (IR) for the mean absolute error (MAE), structural similarity index (SSIM), and peak signal-to-noise ratio (PSNR) obtained for each training paradigm. The total training time required to complete 25 rounds is also reported.}
    \label{tab:train_paradigm_investigation}
\end{table}

The results presented in Table \ref{tab:train_paradigm_investigation} further supported the superiority of the proposed Random Multi-2D approach over the alternative 2D methods evaluated. Specifically, the Random Multi-2D method achieved performance in terms of image similarity metrics comparable to the 2D+ approach, where each model was specifically trained on a single plane (axial, coronal and sagittal). Our training method required only a single model for all anatomical planes, thereby simplifying the federated training and inference processes. Additionally, the training time was significantly reduced with the Random Multi-2D model completing training in 12 hours and 34 minutes, compared to 16 hours for the 2D+ models. In contrast, the 2D patch-based method requiring less training time but exhibited inferior performance across all image similarity metrics. 

In comparison with the classical Multi-2D \cite{Spadea2021} method, the proposed Random Multi-2D approach yielded superior image similarity metrics, albeit at a higher computational cost, which can be attributed to the shuffling operation conducted by local clients during federated training. Moreover, the utilisation of our method enabled the federated model to attain superior outcomes in comparison to alternative training methods after only 5 rounds.

The findings of this additional study demonstrated that the Random Multi-2D paradigm not only exhibited competitive performance in comparison to the 2D+ approach, but also demonstrated improved computational efficiency and substituting a more generic model for three specific models. Furthermore, the proposed training method yielded superior generalisation performance compared to the classical Multi-2D method, thus justifying the suitability of the Random Multi-2D paradigm for FL MRI-to-sCT scenarios.

\section{Centre B artefact measurement}
\label{app:centreb_artefact}
\begin{figure}[H]
    \centering
    \includegraphics[width = 1\linewidth]{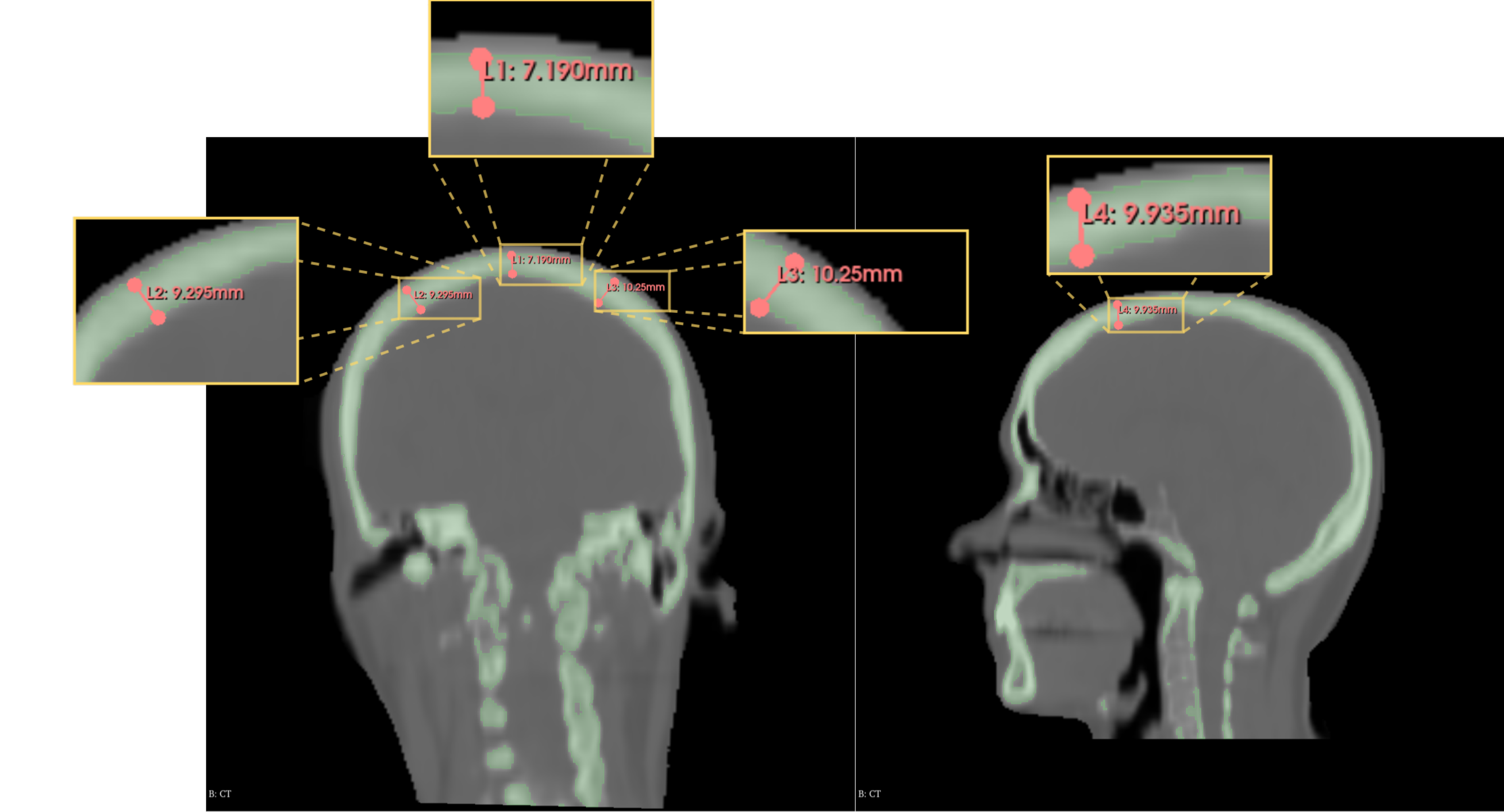}
    \caption{Measurement of a portion of the skull to highlight the partial volume artefact due to the higher slice thickness in Centre B CTs.}
    \label{fig:CentreBArtefact}
\end{figure}

\section{Aggregation strategies investigation}
\subsection*{Convergence trends}
\label{app:aggregation_trends}
\begin{figure}[H]
    \centering
    \includegraphics[width = 1\linewidth]{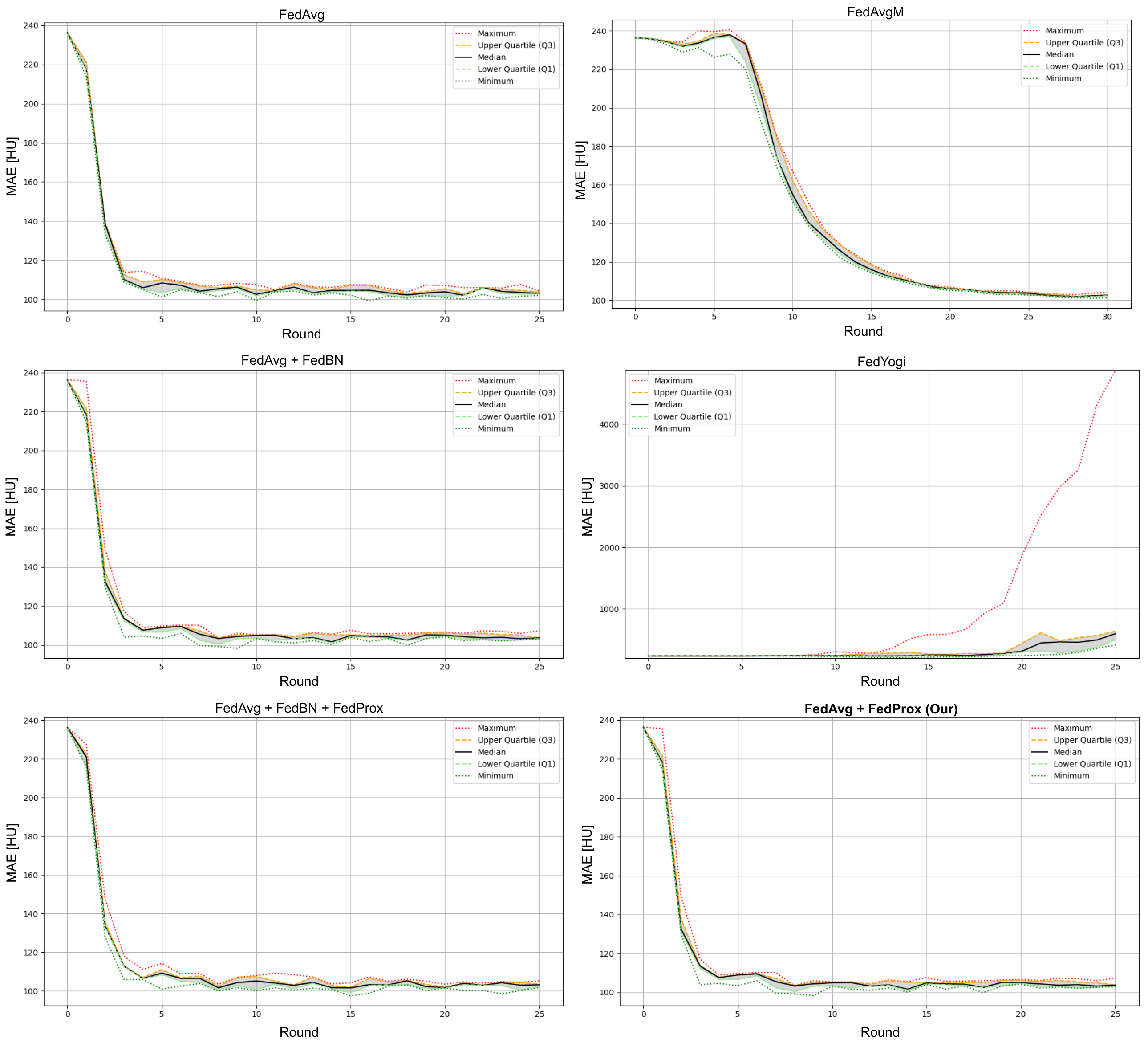}
    \caption{Comparison of convergence trends of aggregation strategies.}
    \label{fig:AggregationsPlots}
\end{figure}

\subsection*{Generated sCTs}
\label{app:aggregation_sCTs}
\begin{figure}[H]
    \centering
    \includegraphics[width = 1\linewidth]{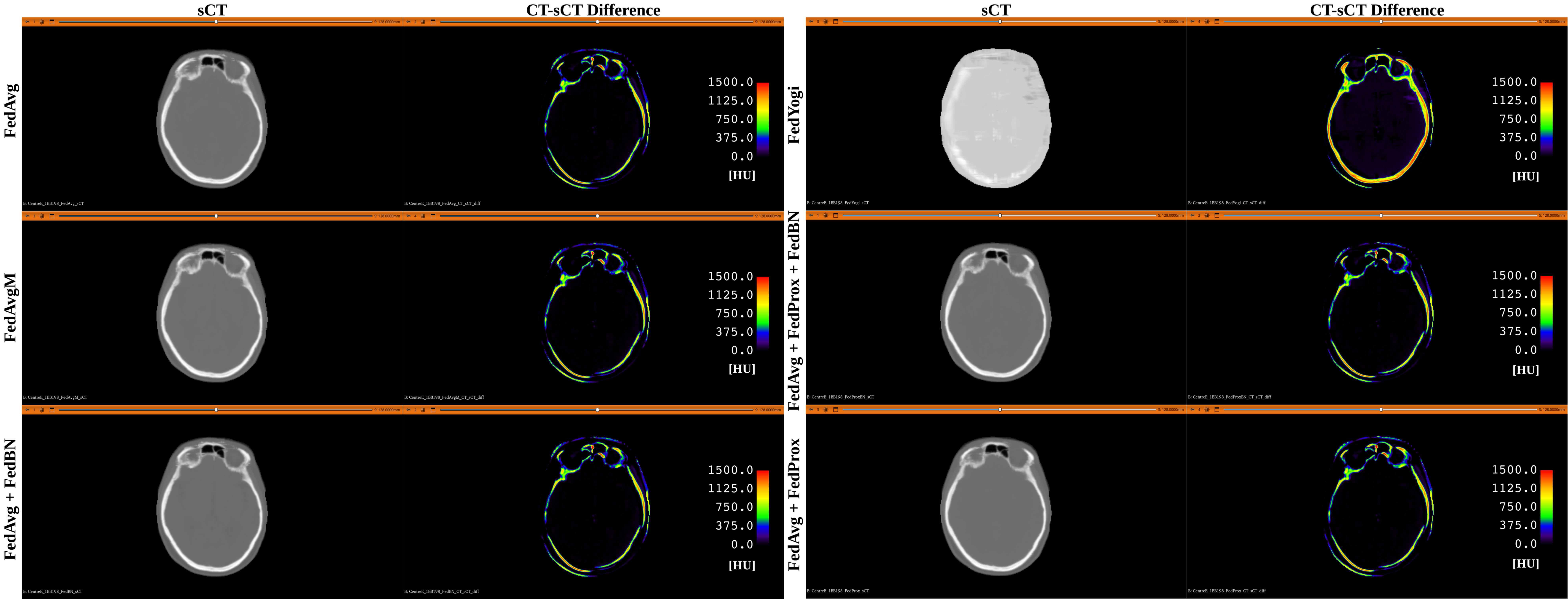}
    \caption{Visual comparison of sCT generated with different aggregation strategies on an axial slice from a patient of Centre E. The discrepancy between the ground truth CT and the sCT is highlighted on the right.}
    \label{fig:AggregationsVisualResults}
\end{figure}

As shown in Table \ref{tab:strategies_comparison}, the evaluated strategies produced qualitatively comparable results, with similar mean absolute errors (MAE). This observation is further supported by Figure \ref{fig:AggregationsVisualResults}, which provides a visual comparison of the results obtained using different aggregation strategies on an axial slice of an example sCT from the same patient of the unseen Centre E. The discrepancy between the ground truth CT and the sCT is also illustrated.
Notably, the sCT corresponding to FedYogi demonstrated a more pronounced discrepancy from the ground truth CT, in line with the highest MAE value reported in Table \ref{tab:strategies_comparison}. This finding confirmed the limitations of FedYogi for the MRI-to-sCT task compared to the other aggregation strategies, as well as the pivotal role of the aggregation strategy in the federated context. Indeed, as well as determining the quality of the generated sCT, the strategy had a decisive influence on model convergence.

\bibliographystyle{abbrv}

\end{document}